\documentclass[useAMS,usenatbib,usegraphicx]{mn2e}
\usepackage{amssymb}
\usepackage{amsmath}
\usepackage{natbib}
\usepackage{graphics}
\usepackage{color}
\usepackage{ulem}
\def\apj{ApJ}%
\def\mnras{MNRAS}%

\title[Turbulence-induced magnetic fields in shock precursors]{Turbulence-induced magnetic fields in shock precursors}
\author[M.V. del Valle,  A. Lazarian and R. Santos-Lima]{M.V. del Valle$^{1}$\thanks{E-mail: maria@iar-conicet.gov.ar}, A. Lazarian$^{2}$\thanks{E-mail: alazarian@facstaff.wisc.edu} and R. Santos-Lima$^{3}$\thanks{E-mail: rlima@astro.iag.usp.br}\\
$^{1}$Instituto Argentino de Radioastronom\'{\i}a, C.C.5, (1894) Villa Elisa, Buenos Aires, Argentina\\
$^{2}$Department of Astronomy, University of Wisconsin,
             475 North Charter Street, Madison, WI 53706, USA\\
$^{3}$Instituto de Astronomia, Geof\'isica e Ci\^encias Atmosf\'ericas, Universidade de S\~ao Paulo, S\~ao Paulo, SP 05508-090, Brazil}

\begin{document}

% \date{Accepted 1988 December 15. Received 1988 December 14; in original form 1988 October 11}

% \pagerange{\pageref{firstpage}--\pageref{lastpage}} \pubyear{2002}

\maketitle

\label{firstpage}

\begin{abstract}
{
Galactic cosmic rays are believed to be mostly accelerated at supernova shocks. However, the interstellar magnetic field is too weak to efficiently accelerate galactic cosmic rays up to the highest energies, i.e. $10^{15}$ eV. A stronger magnetic field in the pre-shock region could provide the efficiency required. Bell's cosmic-ray nonresonant streaming instability has been claimed to be responsible for the amplification of precursor magnetic fields. However, an alternative mechanism { has} been proposed in which the cosmic-ray pressure gradient forms the shock precursor and drives turbulence, amplifying the magnetic field via the small-scale dynamo. A key ingredient for the mechanism to operate are the inhomogeneities present in the interstellar medium (ISM). These inhomogeneities are the consequence of turbulence. In this work we explore the magnetic field amplification in different ISM conditions through 3D MHD numerical simulations.
}
\end{abstract}


%%==============================================================================
%%
%%
Cosmic rays (CRs) are high-energy particles with energies between $10^8-10^{22}$ eV 
whose origin has been subject of debate since the beginning of research in the field 
\citep[e.g.,][]{ginzb}. From energetic constraints it can be  shown that CRs with 
energies up to $10^{15}$ eV (the so called {\it knee}) are of galactic origin 
(see Blasi 2013, and references there in). The main  galactic CR sources are believed 
to be supernova remnants (SNRs) shocks, where particles are accelerated through the 
diffusive shock mechanism \citep[e.g.,][]{Bell,drury84}. 

The pressure of CRs can be comparable to the pressure of the background medium, 
so they produce dynamical effects on the background plasma and magnetic field 
\citep[e.g.,][]{Axford82}; CRs modify shocks \citep[see,][and references there in]{malkov2001,blasi2013}. The accelerated particles can extract significant energy from the shocked flow. The high-energy particles diffuse ahead of the shock producing a pressure gradient upstream of the shock transition that smoothly decelerates and compresses flow into the shock, and a shock precursor is formed.   

A  serious problem on the theory of CR acceleration in SNR shocks is that the magnetic field strengths in the upstream region, i.e. the $\sim$ 5 $\mu$G ISM magnetic fields, are too weak for efficiently accelerate CRs of PeV energies \citep{lagage83}. PeV CRs have long mean free paths in such fields and have a high probability of escaping the shock, so they are not subject to further acceleration. The problem might be solved if the magnetic field in the pre-shock region can be much stronger than its interstellar value and if the free energy available for the shock is sufficient to generate much larger fields. 

Furthermore,  indirect observational arguments  require  higher magnetic fields at the SNR shocks, higher than expected by simple shock compression. Detection of rapid variability in X-ray emission from a supernova remnant indicates an amplification factor of the magnetic field of more than 100 \citep{uchiyama09}. Upstream magnetic field amplification not only should occur in  SNR shocks, but in other strong shock waves  in the ISM produced, for example, by massive star winds, stellar outflows, jets, etc. Also observational evidence suggests that radio relics host relatively large magnetic fields \citep{markevitch05,vanWeeren11}. Magnetic field amplification might occur at the weak shocks  that produce the radio relics \citep{bruggen13}.

\citet{Bell04} proposed the current driven instability to overcome this problem. 
The driving electric current of the instability comes from drift (streaming) 
of the escaping CRs. The reacting current of the background plasma leads to a transverse force that can amplify transverse perturbations in the magnetic field. 
However, this mechanism can only generate fields on scales smaller than the gyro-radius of the driving particles, and the generated fields are  on scales too small to be used to accelerate the highest energy particles  \citep{schure12,beresnyak14}. Other process might occur simultaneously. Another instability has been proposed \citep[e.g.,][]{Diamond2007, malkov09}, the Drury acoustic instability \citep{drury84}, which is the enhancement of compressible perturbations by the CR pressure gradient. This instability potentially can have faster growth rates 
than the Bell instability and  operates on larger scales. The relative role of different instabilities in generating magnetic field in front of the shock is currently the field of active research. 

\citet{Beresnyak09} (henceforth  BJL09) have presented an alternative process that 
can provide fast magnetic field generation. In the BJL09 model the magnetic field is generated by purely non-linear fluid mechanisms. The CR pressure gradient accelerates differentially the denser and lighter parts of the ISM upstream flow carrying density inhomogeneities, generating in this way local shear or vorticity. This vorticity drives a (supersonic) turbulence cascade. The magnetic energy is amplified on scales of the order and below the scales of the vorticity structures (of the order of the density structures scales) through the small-scale dynamo \citep[e.g.,][]{cho09,beresnyak14}. This idea was later elaborated by \citet{drury12}.

The inhomogeneities observed in the ISM covering an extended range of scales, the spectrum of these density fluctuations are consistent with  Kolmogorov turbulence, the so-called “Big Power Law in the Sky” \citep{armstrong95,chepurnov10}; density inhomogeneities are  present in many astrophysical plasmas and are usually associated with magnetohydrodynamic (MHD) turbulence \citep[e.g.,][]{elmegr2004,laz_oph2009}. The ISM is composed by different phases according to the physical conditions of the material. These physical conditions establish different 
regimes of MHD turbulence. The typical values for the density fluctuations amplitudes and scales from the turbulent ISM would allow a magnetic field amplification  larger in scales and strength than that provided by the Bell's instability (Beresnyak, Jones, \& Lazarian 2009). 

\citet{drury12} presented the first study on the turbulent magnetic field amplification driven by CR pressure gradients using 2D MHD simulations; later on they extended this work and compared the results with 3D simulations \citep{downes14}. 
They found that independent of the detailed plasma physics, as long as there is a 
cosmic ray precursor with a significant associated pressure gradient, and assuming the inflowing medium is inhomogeneous, the magnetic field can be amplified in the precursor by a factor $\sim 5-20$ (depending mainly on the direction of the uniform initial magnetic field). 

\citet{bruggen13} also studied upstream magnetic field amplification by turbulence using 3D and mostly 2D MHD simulations; weak shocks of galaxy clusters were also considered. However, due to the amplification relying on the hydromagnetic turbulence the 2D treatment  is not adequate (because the physics of the turbulent cascade is different from the 3D real case). In addition, the above mentioned works considered a synthetically built density fluctuations spectrum for representing the medium inhomogeneities.

In this work we also study turbulence-induced magnetic field amplification in shock upstream, exploring the details of the turbulence developed in the precursor. We use density structures obtained in different regimes of compressible MHD isothermal 
turbulence simulations to represent the inhomogeneities in the medium swept by the shock, in order to compare with the simplest density structures built artificially.
We aim at studying the magnetic field amplification for different conditions of the ISM, under the ideas presented by BJL09. Here we deal only with magnetic field amplification by turbulence in the pre-shock region. Post-shock turbulence can further amplify the magnetic field downstream; this situation is studied elsewhere \citep[e.g.,][]{giacalone07,gou12,mizuno14}.

This paper is organized as follows. In the next section we describe the magnetic 
field amplification model developed in BJL09 and we highlight the existence of the density homogeneities in the ISM, one of the main ingredients of the mechanism.
Section~\ref{sec:numset} describes the numerical setup we adopt in the MHD simulations. In the Section~\ref{sec:results} we present the results and data analysis. An extended  discussion in made in Section~\ref{sec:discussion}. 
Finally, in Section~\ref{sec:conclusions}, we draw our conclusions.

%%==============================================================================
%%
%%
\section{Magnetic field amplification mechanism}
\label{sec:model}

Here we describe briefly the main ideas in the model developed in BJL09  for magnetic field amplification. For further details and argumentations the reader is referred to the original paper.  

\begin{figure}
 \centering
\includegraphics[width=0.8\linewidth, trim=0cm 2cm 0cm 2cm, clip=true]{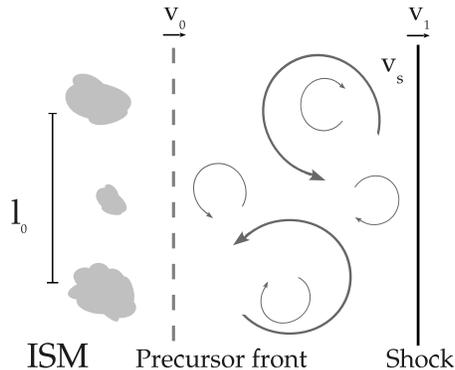}
 \caption{2D scheme of the solenoidal motions excited by cosmic-ray pressure in the shock precursor.}  
\label{fig:solenoidal}
\end{figure}

The fluid elements passing through the precursor have density inhomogeneities. As already mentioned, some level of density inhomogeneity in astrophysical fluids is naturally generated by the combination of compressible turbulence, gravity and cooling \citep{armstrong95,burkhart09,federrath10,chepurnov10}. When this inhomogeneous plasma flows into the precursor with speed $v_{0}$ (the shock front velocity in the reference frame of the upstream flow), it is decelerated by the CR pressure gradient down to a velocity $v_1$ at the shock front. This situation is illustrated in Fig.~\ref{fig:solenoidal}. The interaction between the fluid and the CR pressure creates perturbations of velocity $\sim (v_0 - v_1)$, as a consequence of the differences between the ballistic velocity that highest density regions have and the full deceleration suffered by  the  lowest density regions. The resulting velocity field is partially solenoidal due to the shear caused by the different velocities acquired by the light and dense parts of the gas.

The amount of solenoidal velocity can be parametrized as
\begin{equation}
v_s = A_s\,(v_0 - v_1),
\end{equation}
with $A_s \leq 1$.  The characteristic scale of the solenoidal perturbations $l$ is, for the case of supersonic  ISM, the distance between slow sub-shocks (see Section~\ref{sec:ism}), which is smaller than the turbulence outer scale $l_0$. $l$ can be of the order of parsecs (see discussion in the next subsection). These solenoidal velocities develop a turbulent cascade, which amplify the weak magnetic magnetic field embedded in the precursor through the turbulent Small-Scale Dynamo (SSD).

The SSD has three main stages: the kinematic stage during which magnetic energy grows exponentially; the linear stage and a saturation stage \citep[e.g.,][]{cho09,beresnyak14}. The kinematic stage in astrophysical problems is not relevant because its saturation time-scale is very short compared to astrophysical time-scales. Therefore, it is  always assumed that the kinematic dynamo is saturated and the dynamo is in the linear stage. In this stage the magnetic energy increases  linearly with time, i.e., 
\begin{equation}\label{rate}
\frac{1}{8\pi}\frac{dB^2}{dt} = A_d \epsilon, 
\end{equation}
here $\epsilon$ is the energy transfer rate of the turbulence and $A_d$ can be thought as the efficiency of the SSD. The energy transfer rate can be estimated as $\epsilon=\rho v_s^3/l$. 

At some time the magnetic field reaches equipartition with the turbulent velocity field for motions at the scale $l^*$.  This scale grows with time. On scales smaller than $l^*$ magnetic and velocity perturbations form a MHD turbulent cascade with a  steep spectrum. On scales larger than $l^*$, the magnetic field has a shallow spectrum and the velocity has a Kolmogorov spectrum.

When $l^*$ approaches $l$, the SSD enters the saturation stage in which the magnetic field grows more slowly than in the previous linear stage. However, in the physical scenario considered here, the saturation stage is never reached, since limited time is available for field amplification.

The mechanism efficiency depends on the relative amplitude of the initial density perturbations, i.e. $\delta \rho / \rho$. When  $\delta \rho / \rho$ $\sim$ 1, the mechanism is expected to be more efficient. Even when the ratio $\delta \rho / \rho$ is initially small, the initial density perturbations could be enhanced in the precursor up to the order of unity by different mechanisms. Under some conditions Rayleigh-Taylor-like  instabilities can strongly enhance turbulence in the CR precursor. 

%%==============================================================================
%%
%%
\subsection{Turbulence in the interstellar medium}
\label{sec:ism}

Many possible sources of turbulence in the ISM exist operating at different scales and injecting different powers. The ISM is classified in phases, according to the temperature and state of the Hydrogen, of which it is almost fully composed. Cold regions in which hydrogen is in molecular form include dark clouds and molecular clouds, and in which hydrogen is atomic include cold neutral medium, photo-dissociation regions, reflection nebula, etc. The warm ISM includes warm neutral medium of atomic hydrogen and ionized regions such as the so-called warm ionized medium, and H II regions. The hot ISM includes coronal gas, hot ionized medium, etc. \citep[e.g.,][]{Draine98}. The physical conditions of the ISM phases, including the degree of magnetization, stablish the sonic and Alfv\'enic regime of the turbulence. 
   
The turbulence in hot and warm ISM is usually subsonic or trans-sonic with respect to the ambient turbulent motions \citep[see][]{2008ApJ...686..363H,2012ApJ...749..145B}. A generic property of the cold ISM is cooling which keeps temperatures  low, making ISM turbulence highly compressive. Thus the Mach numbers increase in cold HI \citep[see][]{chepurnov10b}. In dense regions, observations reveal that turbulence in molecular clouds and star forming regions can be highly supersonic \citep[see][]{crutcher99,falgarone08}. In general, the warm ISM typical observed Mach numbers are  between 1 and 10. The cold and dense ISM has   Mach numbers between 10 and 50 \citep[e.g.,][]{mac04}. 

%%==============================================================================
%%
%%
\section{Numerical setup}
\label{sec:numset}

In order to study the magnetic field amplification on the shock precursor we 
perform MHD simulations of a supersonic flow in the shock referential frame. 
We employ the PLUTO code \citep{mignone07}, a conservative shock-capturing code, 
for solving the dimensionless ideal MHD equations:

\begin{eqnarray}
\frac{\partial \rho}{\partial t} + \mathbf{\nabla} \cdot \left(\rho
		      \mathbf{v}\right)  & = & 0 \label{mass} \\
\frac{\partial \rho \mathbf{v}}{\partial t} + \nabla\cdot\left( \rho \mathbf{v} 
	\mathbf{v} + P\right) + \mathbf{B}\times(\nabla\times\mathbf{B}) & = & 
\mathbf{F}_{\rm CR}, \label{neutral_mom} \\
\frac{\partial e}{\partial t} + \mathbf{\nabla} \cdot \left[ \left(e +
		P + \frac{B^2}{2}\right)\mathbf{v}-\mathbf{B}(\mathbf{v}\cdot\mathbf{B})\right] & = &
\mathbf{F}_{\rm CR}\cdot\mathbf{v}\\
\frac{\partial \mathbf{B}}{\partial t} - \mathbf{\nabla}\times (\mathbf{v}\times \mathbf{B}) & = & 0, \label{B_eqn} \\
								  \nabla\cdot\mathbf{B} & = & 0 \label{divB}
\end{eqnarray}
where $\rho$ is the density, $\mathbf{v}$ is the velocity, $P$ is the thermal pressure, $\mathbf{B}$ is the magnetic field, $e = P/(\gamma - 1) + \rho v^2/2 + B^2/2$ the total energy density ({ excluding the CR energy}), $\gamma$ is the adiabatic index, and $\mathbf{F}_{\rm CR}$ is the force per volume due to the CR pressure gradient (see below). We use the third order Runge-Kutta scheme for time evolution, and the fluxes are calculated employing the HLL solver and parabolic interpolation. The magnetic divergence is controlled using a cleaning divergence scheme. The MHD equations are solved in a three-dimensional Cartesian box discretized by a uniform grid.

The CR pressure is directly related to the flow kinetic energy available in the pre-shock (at the shock referential). In our simulations the shock front is considered flat and localized at the right boundary of the computational domain. We model the force due to the CR pressure $\mathbf{F}_{\rm CR}$ as decreasing exponentially with 
the distance from the shock position $x_{\rm s}$\footnote{We are considering a constant diffusion scale (diffusion coefficient), that attributed to the highest cosmic-ray energy; hence such a spatial exponential decay is naturally expected.}, through the equation:

\begin{equation}\label{force}
\mathbf{F}_{\rm CR}  =  - \frac{\eta}{D_{\rm CR}} \left\langle p_x(x_{\rm s}) \right\rangle \left\langle v_x(x_{\rm s}) \right\rangle e^{-|x-x_{\rm s}|/D_{\rm CR}} \, \mathbf{i}.
\end{equation}

Here  $\left\langle p_x(x_{\rm s}) \right\rangle_{yz} = \langle \rho(x_{\rm s}) v_x(x_{\rm s}) \rangle_{yz}$ and $\left\langle v_x(x_{\rm s}) \right\rangle_{yz}$ are the averages in the $yz$-plane of the $x$-component of the momentum and of the $x$-component of the velocity respectively; $\eta$ is an efficiency parameter smaller than 1. Following \citet{drury12} \citep[see also][]{vink12}, we adopt $\eta = 0.6$.
The CR diffusion scale $D_{\rm CR} = \kappa_{\rm CR} / v_0$, where $\kappa_{\rm CR}$ is the CR diffusion coefficient in the preshock region and $v_0$ is the preshock flow velocity, is considered constant in the entire domain and in time. The computational domain, in the direction parallel to the shock propagation ($x$-direction), has length $L = D_{\rm CR}$.

We apply the physical constraint that the force produced by the CR pressure is halted if the total Mach number~\footnote{In the models description the  parameter ``${\cal M}_{\rm s}$'' refers to the value of the Mach number of the pre-shock flow at infinity (far from the shock); in the simulations it is the physical value at the inflow boundary.} ${\cal M}_{\rm T} = \langle v_x(x_{\rm s}) \rangle_{yz} / \sqrt{\langle c_s(x_{\rm s}) \rangle_{yz}^2 + \langle v_{\rm A}(x_{\rm s}) \rangle_{yz}^2}$, where $v_{\rm A} = B/\sqrt{4 \pi \rho}$ is the local Alfv\'en velocity, becomes less than unity close to the shock (although this condition is never fulfilled in our simulations). 

It should be remarked that we do not use explicit viscosity and/or resistivity in our calculations in order to maximize the Reynolds and magnetic Reynolds numbers, which are very high in astrophysical flows~\footnote{Due to the non-vanishing numerical resistivity, MHD simulated flows have an upper limit Reynolds number. This upper value is still much lower than the real  astrophysical Reynolds and magnetic Reynolds numbers.}. We do not include heat conduction and/or cooling in our simulations.

%%%%%%%%%%%%%%%%%%%%%%%%%%%%%%%%%%%%%%%%%%%%%%
\subsection{Initial conditions}
\label{sec:inicon}

The domain is initially filled by a density field $\rho$ composed by a uniform component plus an inhomogeneous, ``turbulent'' one: $\rho = \rho_0 + \delta \rho_{\rm turb}$, with $\langle \rho \rangle = \rho_{0} = 1$ (in code units), where 
the brackets $\langle \cdot \rangle$ denote a spatial  average over the domain. 
The initial pressure is uniform and given by $P = c_s^2 \rho_0 / \gamma$, where $c_s = 1$ (in code units) is approximately the initial average sound speed, and we use the adiabatic index $\gamma=5/3$ in all our simulations. The initial flow has a uniform velocity $v_{0} = 100 c_s$ in the $x$-direction. The initial magnetic field is uniform in the $y$ direction, i.e. a perpendicular shock, with $B_0 = 0.1$. 

We define the initial sonic Mach number and Alfv\'enic Mach number in the preshock 
respectively as ${\cal M}_s \equiv v_{0} / c_s$ and ${\cal M}_{\rm A} \equiv v_{0} / B_0$. Therefore, through this study we use ${\cal M}_{\rm s} = 100$ and ${\cal M}_{\rm A} = 1000$, unless stated otherwise. 

We parametrize the ``turbulent'' density fluctuations $\delta \rho_{\rm turb}$ used in the initial condition by two dimensionless quantities: the ratio between the typical length of the density structures and the CR diffusion scale $\lambda_{\rho}/D_{\rm CR}$, and the relative amplitude of these density fluctuations $\delta \rho_{rms}/\rho_0$ (the density contrast). 

Table~\ref{tab:amodels} shows the parameters of the models presented in this work. 
The initial density fluctuation fields are calculated in different ways: 
for Models A, a spectrum of plane waves with random phases is imposed (synthetic structures), and for Models B, the density field is taken from a snapshot of MHD 
turbulence simulations. These two different initial conditions are detailed below. 

\begin{table*}
\begin{center}
\caption{Models parameters \label{tab:amodels}}
\begin{tabular}{l l l l l l l}
\hline\hline
Model & 
$\lambda_{\rho} / D_{CR}$ & $\delta \rho_{rms} / \rho_0$ & 
${\cal M}_{s \,{\rm turb}}^*$ & 
${\cal M}_{A \,{\rm turb}}^*$ &
$L \times L_y \times L_z$ $(\div D_{CR})$ &
$N_x \times N_y \times N_z$ \\
\hline\noalign{\smallskip}
AI    & 0.05   & 0.84 & - & - & $1 \times 0.125 \times 0.125$ & $1024 \times 128 \times 128$ \\
AII   & 0.05   & 0.24 & - & - & $1 \times 0.125 \times 0.125$ & $1024 \times 128 \times 128$ \\
%AII   & 0.05   & 0.54 & - & - & $1 \times 0.125 \times 0.125$ & $1024 \times 128 \times 128$ \\
%AIV   & 0.025  & 0.86 & - & - & $1 \times 0.0625 \times 0.0625$ & $2048 \times 128 \times 128$ \\
%AV    & 0.0125 & 0.86 & - & - & $1 \times 0.03125 \times 0.03125$ & $4096 \times 128 \times 128$ \\
AIII  & 0.0125 & 0.84 & - & - & $1 \times 0.125 \times 0.125$ & $2048 \times 256 \times 256$ \\
HAI   & 0.05   & 0.85 & - & - & $1 \times 0.125 \times 0.125$ & $2048 \times 256 \times 256$ \\
\hline
BI   & 0.05 & 0.90 & 2.3 & 2.9 & $1 \times 0.125 \times 0.125$ & $1024 \times 128 \times 128$ \\
BII  & 0.05 & 0.28 & 0.7 & 2.7 & $1 \times 0.125 \times 0.125$ & $1024 \times 128 \times 128$ \\
BIII & 0.05 & 0.91 & 2.3 & 0.6 & $1 \times 0.125 \times 0.125$ & $1024 \times 128 \times 128$ \\
BIV  & 0.05 & 0.34 & 0.7 & 0.7 & $1 \times 0.125 \times 0.125$ & $1024 \times 128 \times 128$ \\
\hline\hline
\end{tabular}
\end{center}
\end{table*}

%%%%%%%%%%%%%%%%%%%%%%%%%%%%%%%%%%%%%%%%%%%%%%
\subsubsection{Synthetic density fluctuations}

Using a density field built from a composition of random-phased wave modes with an imposed spectrum allows us 
to directly control the key parameters $\lambda_{\rho}$ and $\delta \rho_{rms}$ of the density structures. According to the BJL09 theory, both  quantities should impact on the magnetic energy growth due to different reasons: the highest the density contrast $\delta \rho_{rms}/\rho_0$ the more intense vorticity the CR pressure generates in the flow, and the smaller the density structures size $\lambda_{\rho}$, the smaller the turn-over times of the largest eddies of the induced turbulence, which means that the turbulence cascade has more time to develop and to amplify the magnetic field before the plasma crosses the shock.

The initial density field is a log-normal distribution given by
\begin{equation}
\rho(x,y,z) \propto \rho_0 \exp(C \delta f(x,y,z))
\end{equation}
where $\delta f(x,y,z)$ is generated by summing over a larger number of discrete modes with the wave-vectors norm $k$ between $k_{\min} = L/\lambda_{\rho}$ (the wave-number characterizing the density fluctuations scale) and $k_{\max} = 2 L_y/N_y$ (the largest wave-number in the box). These modes have random phases and their amplitudes are calculated so that the final unidimensional power spectrum of $\delta f$ follows the power law $P(k) \propto k^{-5/3}$. The amplitudes of the modes with $k$ smaller than $k_{\min}$ are strongly damped. The field $\delta f$ is normalized in order to approximate the density variance to the desired value. Finally, the density field is also re-normalized in order to guarantee $\langle \rho \rangle = \rho_0$. The values $\delta \rho_{rms}/\rho_0$ shown in Table 1 are calculated from the resulting density field. 

\begin{figure}
\begin{tabular}{c}
\includegraphics[width=.7\linewidth, trim= 0cm 0cm 0cm 0cm, clip=true,angle=270]{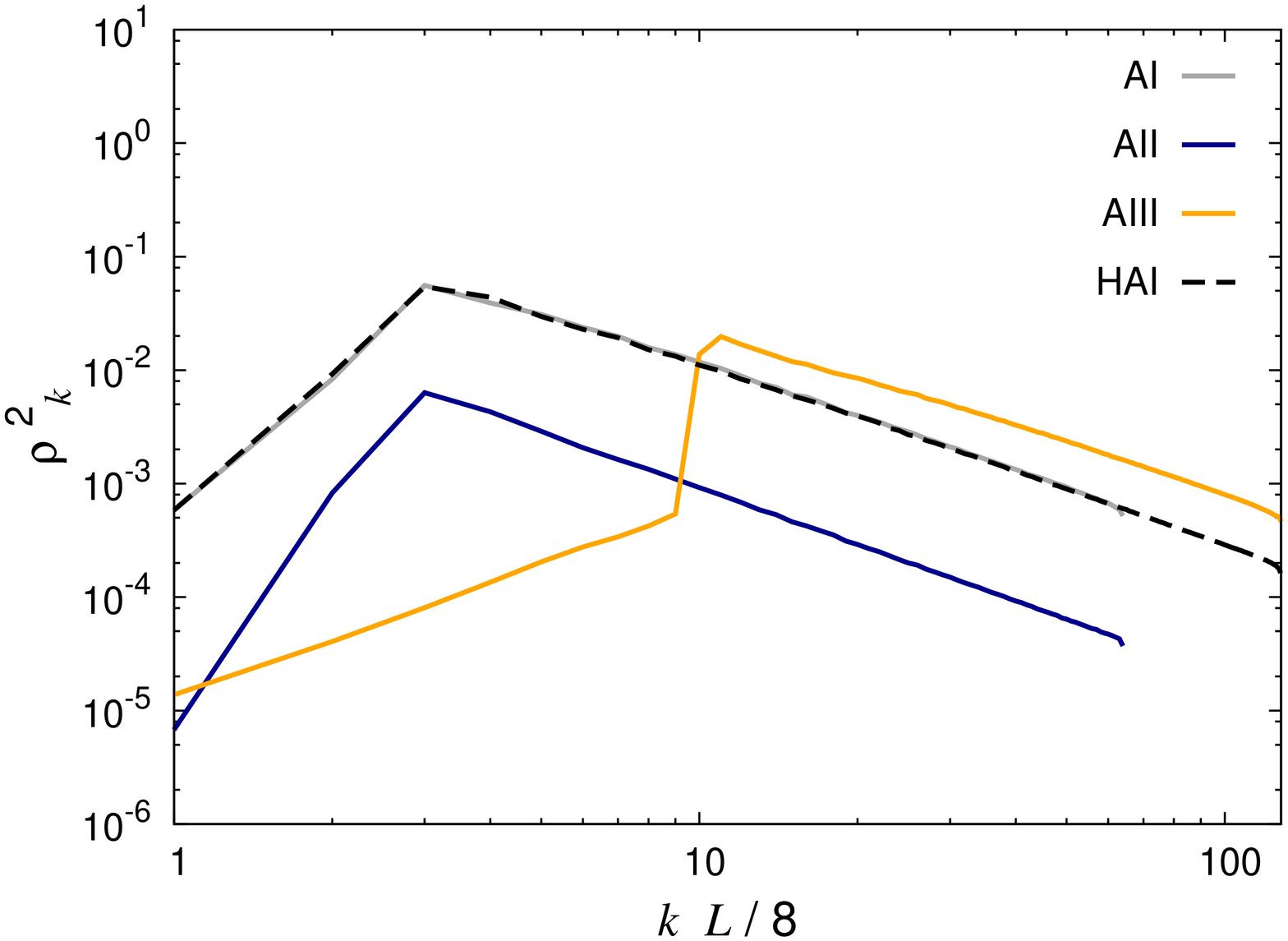}\\
\includegraphics[width=.7\linewidth, trim= 0cm 0cm 0cm 0cm, clip=true,angle=270]{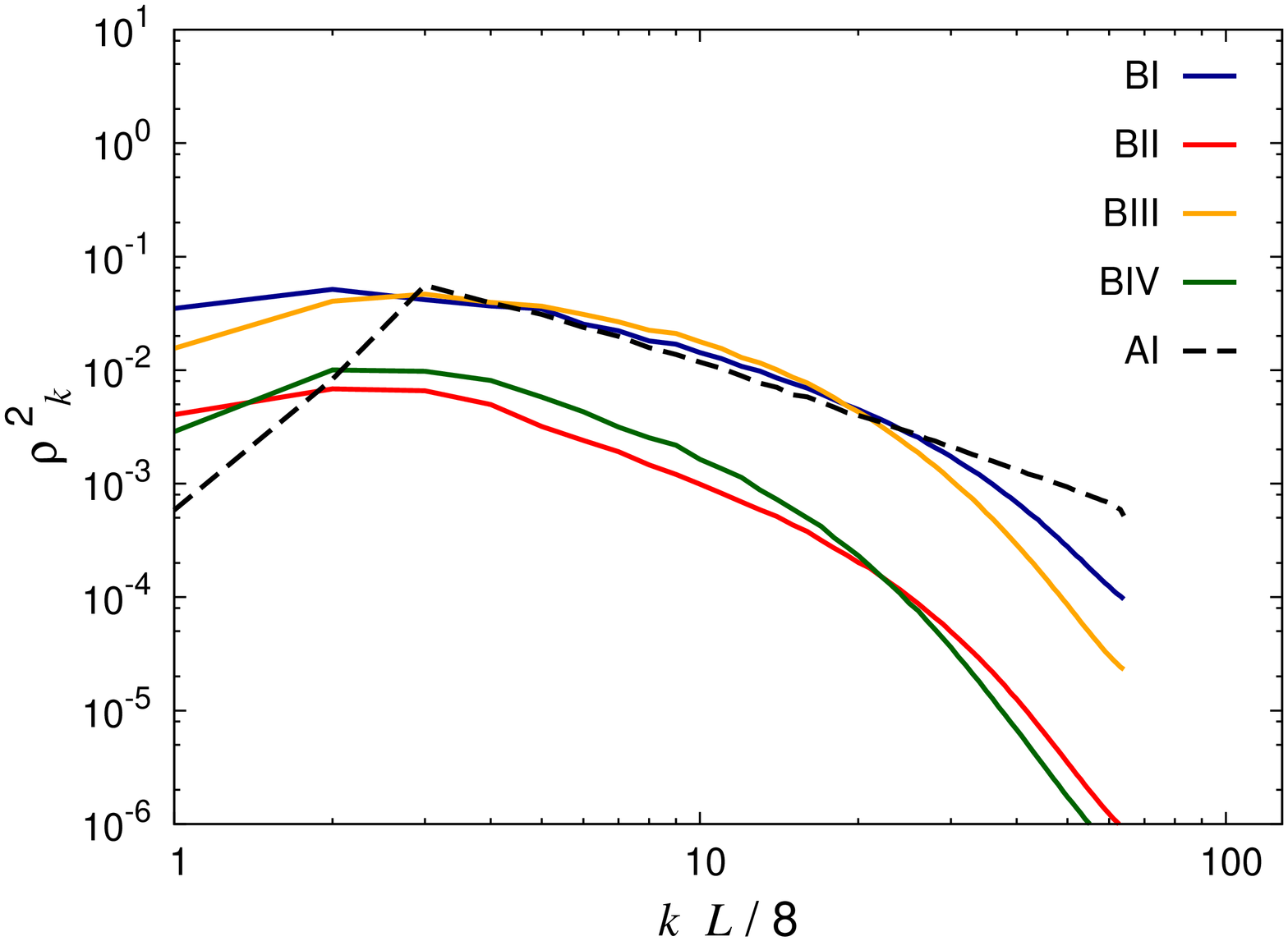}
\end{tabular}
\caption{Power spectrum of the initial density field $\rho$
for Models A ({\it top}) and Models B ({\it bottom}) presented in Table 1.}
\label{fig:ps_initial-rho}
\end{figure}

\begin{figure*}
\begin{center}
\begin{tabular}{c}
{\includegraphics[width=.13\linewidth, trim= 0.62cm 1.3cm 1.3cm 0cm, clip=true,angle=270]{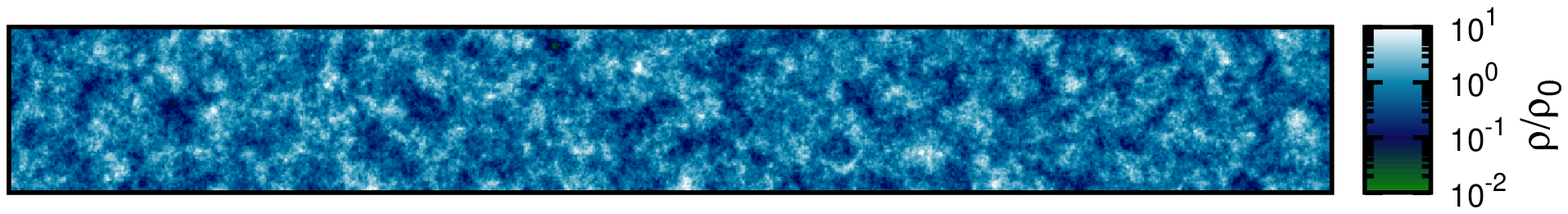}}\\
{\includegraphics[width=.13\linewidth, trim= 0.62cm 1.3cm 1.3cm 0cm, clip=true,angle=270]{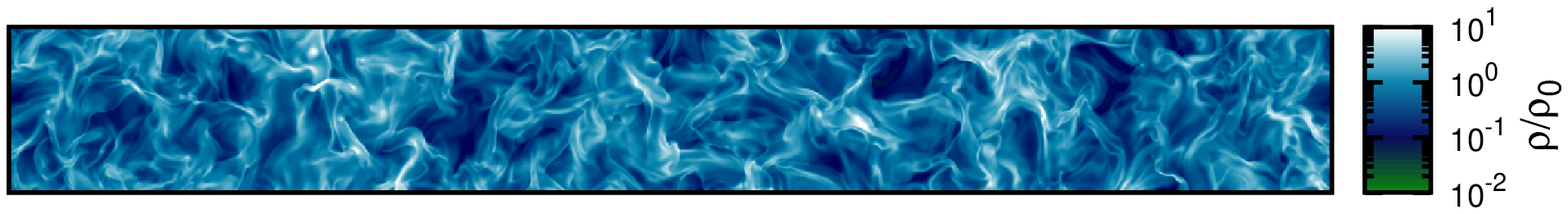}}
\end{tabular}
\end{center}
\caption{Initial density distribution in a central cut of the $xy$-plane of the computational box 
for Model AI ({upper panel}) and for Model BI ({bottom panel}). The parameters 
of the models are listed in Table 1.}
\label{map-initial-rho}
\end{figure*}

Figure~\ref{fig:ps_initial-rho} (top) shows the unidimensional power spectrum of the initial density fields for the models A presented in Table \ref{tab:amodels}. The upper map in Figure~\ref{map-initial-rho} shows the initial density distribution in a central cut of the $xy$-plane of the box for model AI.

\begin{figure*}
\begin{center}
\begin{tabular}{c}
{\includegraphics[width=.13\linewidth, trim= 0.62cm 1.3cm 1.3cm 0cm, clip=true,angle=270]{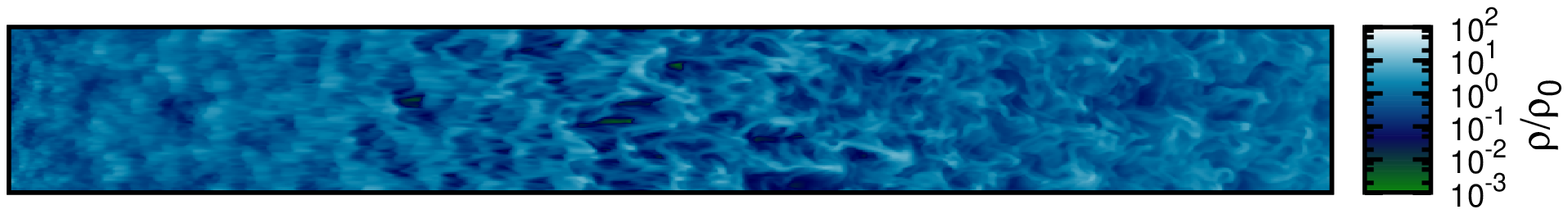}}\\
{\includegraphics[width=.13\linewidth, trim= 0.62cm 1.3cm 1.3cm 0cm, clip=true,angle=270]{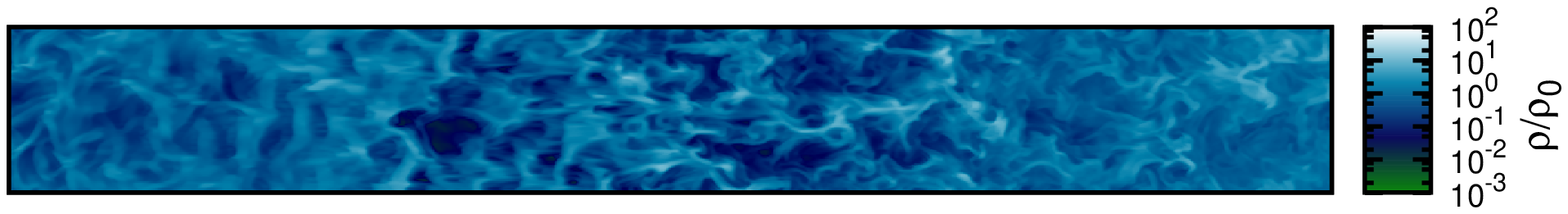}}\\
\end{tabular}
\end{center}
\caption{Final density distribution in a central cut of the $xy$-plane of the computational box 
for Model AI ({upper panel}) and for Model BI ({bottom panel}). The parameters 
of the models are listed in Table 1.}
\label{map-final-rho}
\end{figure*}

\begin{figure*}
\begin{center}
\begin{tabular}{c}
{\includegraphics[width=.13\linewidth, trim= 0.62cm 1.3cm 1.3cm 0cm, clip=true,angle=270]{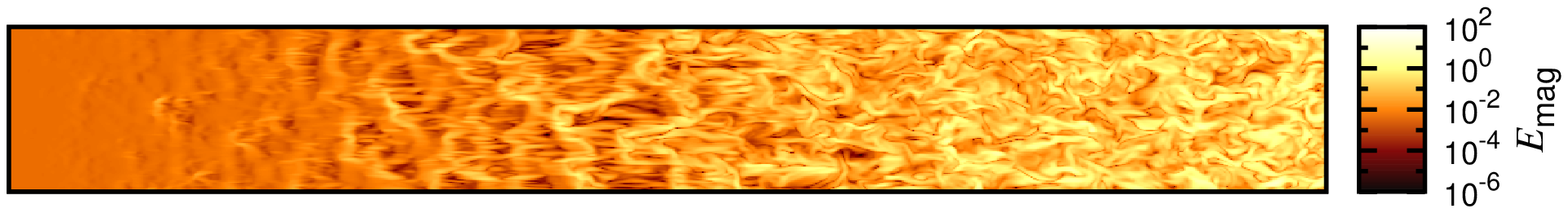}}\\
{\includegraphics[width=.13\linewidth, trim= 0.62cm 1.3cm 1.3cm 0cm, clip=true,angle=270]{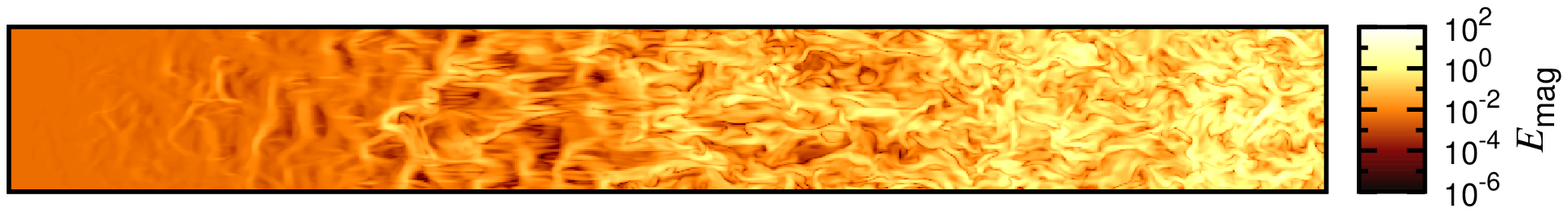}}
\end{tabular}
\end{center}
\caption{Final distribution of the magnetic energy in a central cut of the $xy$-plane of the computational box 
for Model AI ({upper panel}) and for Model BI ({bottom panel}). The parameters 
of the models are listed in Table 1.}
\label{map-final-emag}
\end{figure*}

%%%%%%%%%%%%%%%%%%%%%%%%%%%%%%%%%%%%%%%%%%%%%%
\subsubsection{MHD turbulence density fluctuations}
\label{sec:data}

Density structures in the ISM are generated by the highly compressible MHD turbulence combined with cooling and gravity effects \citep[e.g.,][]{burkhart09}. Therefore, the density fluctuations generated by MHD turbulence represent more realistically the ISM density structures than the synthetic structures described above. It is important to compare then the system evolution of the turbulent density  interacting with the CR precursor when it is artificially produced with that generated by MHD turbulence. With this aim, we also run models where the initial density structure is taken from 
data of MHD turbulence simulations (Models B in Table~\ref{tab:amodels}).

The data are 3D numerical experiments of compressible MHD turbulence for a range of Mach numbers listed in Table~\ref{tab:amodels}. The sonic and Alfv\'enic Mach numbers are defined respectively as 
${\cal M}_{s\,{\rm turb}}^* = \langle |{\bf v}|/c_s \rangle$ and
${\cal M}_{A\,{\rm turb}}^* = \langle |{\bf v}|/v_A \rangle$, where the averaging is taken over the whole volume of the turbulence data. Both sub-Alfv\'{e}nic and super-Alfv\'{e}nic turbulence were considered. We produce these simulations with the appropriate geometry for our study (rectangular), using the {\it AMUN} code\footnote{$http://www.amuncode.org/$} \citep{kowal07}. In these simulations, turbulence is driven solenoidally in a rectangular domain, using an isothermal equation of state. The snapshots used here correspond to an arbitrary time in which the turbulent cascade is already fully developed.

The initial density distribution is illustrated in the lower map in Figure~\ref{map-initial-rho}, which shows the central $xy$-plane of the box for model BI of Table~\ref{tab:amodels}. Figure~\ref{fig:ps_initial-rho} (bottom panel) shows the unidimensional power spectrum of the initial density fields for the models B presented in Table 1. 

For simplicity and to allow an easier comparison with Models A and with previous works, for the initial conditions the physical fields other than density 
are not taken from the MHD turbulence simulation data cubes. 

%%%%%%%%%%%%%%%%%%%%%%%%%%%%%%%%%%%%%%%%%%%%%%
\subsubsection{Parameters choice}

The fact that we keep  the CR diffusion scale $D_{\rm CR}$ constant, which is the only physical length of the problem, allow us to parametrize the system by basically two dimensionless quantities: the density contrast $\delta \rho_{rms}/\rho_0$ and the relative density structures scale $\lambda_{\rho}/D_{\rm CR}$. While the density contrasts can be inferred from observations of the ISM, the density structures size are more difficult to compare with the CR diffusion scale, because this last quantity depends on the CR diffusion coefficient $\kappa_{CR}$, which is very hard to determine  theoretical and observationally.
  
Using for $\kappa_{\rm CR}$ the Bohm diffusion coefficient (as a lower limit) of the most energetic particles assumed to be accelerated at the shock ($E$ $\sim$ $10^{15}$ eV\footnote{This maximum CR energy requires an amplified magnetic field.}), $v_{0} = 3 \times 10^{8}$~cm~s$^{-1}$ and $B \sim 100$~$\mu$G (amplified field), the precursor scale gives $D_{\rm CR} = \kappa_{\rm CR}/v_{0}$ $\geq$  $\kappa_{\rm Bohm}/v_{0}$ $\sim$ $10^{18}$~cm.  

In the warm ISM the density contrast $\delta \rho_{rms}/\rho_0$ are of order of unity \citep{armstrong95}. For smaller Mach numbers $\delta \rho_{rms}/\rho_0$ is expected to decrease, as found from the simulated ISM turbulence (see Table~\ref{tab:amodels}, and \citealt{burkhart09}). The typical outer scales of
the ISM turbulence $L_{\rm Gal}$ extend from 1 pc in the spiral arms,  up to 100 pc in the interarm regions \citep{haverkorn08}\footnote{Note that for the parameters we are considering, turbulence structures present in the precursor are at scales $D_{\rm CR}/80<\lambda_{\rho}<D_{\rm CR}/20$, see values in Table~\ref{tab:amodels}. Therefore the condition $\lambda_{\rho}< L_{\rm Gal}$ is always fulfilled.}.

The range of $\lambda_{\rho}/L$ is chosen in order to minimize the eddy turn-over time of the induced turbulence ($\sim \lambda_{\rho}/v_0$), compared to  the time  the gas takes to cross the precursor length. This length is defined by the computational box length in the $x$-direction, which is the region of influence of the CR pressure gradient: $\sim D_{\rm CR}$. The ratio between the last and the former times gives some estimative of the number of turn-over times of the largest turbulent eddies. Is expected that the larger this ratio ($D_{\rm CR}/\lambda_{\rho}$), the more amplification of the magnetic field is accomplished by the turbulent dynamo before the gas crosses the shock.

Models AI and AII in Table~\ref{tab:amodels} have $D_{\rm CR}/\lambda_{\rho} = L/\lambda_{\rho} = 20$ and different values of the density contrast $\delta \rho_{rms}/\rho_0$.  Model AIII has the same density contrast as model AI, but a larger value of $D_{\rm CR}/\lambda_{\rho}=80$. This model has a different 
resolution in order to avoid  (numerical) dissipative effects to nullify the impact that a smaller value of $\lambda_{\rho}/D_{CR}$ has on the dynamo amplification of the magnetic field. Model HAI has the same parameters as Model AI, except for the higher grid resolution.

The various phases of the ISM exhibit different turbulence regimes, as already mentioned above (see also Section~\ref{sec:ism}). For models B, using density structures from MHD turbulence, we consider: supersonic/super-Alfv\'enic turbulence in model BI, subsonic/super-Alfv\'enic turbulence in model BII, supersonic/sub-Alfv\'enic turbulence in model BIII and subsonic/sub-Alfv\'enic turbulence in model BIV. The density structures produced under these different regimes
are statistically different. 

Our calculations use non-dimensional code units. In order to scale the quantities to physical units, we have to adopt three reference units which are equivalent to the code density, velocity, and length units. Considering typical values for the ISM, the reference physical units can be considered as follows:
$\rho_{0\,{\rm cgs}} = 2.3 \times 10^{-22}$ g~cm$^{-3}$ (initial mean density),
$c_{\rm s0\,cgs} = 10$~km~s$^{-1}$ (initial unperturbed sound speed), 
$D_{\rm CR\,cgs} = 10^{18}$~cm (the CR diffusion scale). In this way, the initial value of the magnetic field is 
$B_{0\,{\rm cgs}} = B_{0} c_{\rm s0\,cgs} \sqrt{4 \pi \rho_{0\,{\rm cgs}}} \approx B_0 \times 50$~$\mu$G, where $B_0$ is the initial magnetic field intensity in code units.

%%%%%%%%%%%%%%%%%%%%%%%%%%%%%%%%%%%%%%%%%%%%%%
\subsection{Boundary conditions}

Except for the $x$-boundary conditions, all boundaries  are set periodic. At the left x-boundary ($x = 0$), the gas inflowing has the same physical fields from the initial conditions, mapped at  $x = - v_0 t$ (and using periodicity in the $x$-direction). At the right $x$-boundary ($x = L$), outflow boundary conditions are used (null gradient across the boundary, and no inflow). However, we keep constant pressure on the right $x$-boundary to avoid sound waves to propagate  inside the domain, as reported in \citet{drury12}. 

%%==============================================================================
%%
%%
\section{Results}\label{sec:results}

\begin{figure}
\begin{center}
\begin{tabular}{c}
\includegraphics[width=.7\linewidth, trim= 0cm 0cm 0cm 0cm, clip=true,angle=270]{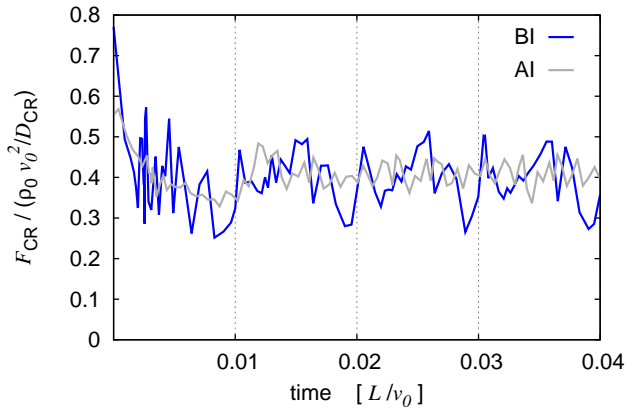} \\
\end{tabular}
\end{center}
\caption{Time evolution of $F_{\rm CR}$ at the right boundary for models AI and BI.}
\label{fig:compAB}
\end{figure}

The system is evolved  during $4L/v_0$ time units. After $t \approx 2L/v_0$, it reaches the statistically stationary state. The value of the force as a function of time is shown in Figure~\ref{fig:compAB} for  models AI and BI. For all the models it is nearly identical, given approximately by $f_{\rm CR} = 0.4 \rho_0 v_0^2/D_{\rm CR}$ after the system stabilization.

We study the profiles of the magnetic field energy, density, and velocity, along the direction parallel to the shock propagation ($x$-axis), and the power spectrum of these fields. The power spectrum are computed in the plane parallel to the shock front (the $yz$-plane) at the last computational cells in the $x$-direction, that is, at the point closest to the shock (unless stated otherwise). These quantities are calculated for each snapshot from $t = 3L/v_0$ to $t = 4L/v_0$ (taken between fixed intervals $\Delta t = 0.2 L/v_0$) and then averaged.

%%%%%%%%%%%%%%%%%%%%%%%%%%%%%%%%%%%%%%%%%%%%%%
\subsection{Profiles} 

As explained in Section~\ref{sec:model}, according to BJL09 the interaction of density fluctuations with the CR precursor produces further turbulence partially solenoidal (that is, vorticity) which amplifies the magnetic field through the small-scale dynamo. 

\begin{figure}
\centering
\includegraphics[width=.7\linewidth, trim= 0cm 0cm 0cm 0cm, clip=true,angle=270]{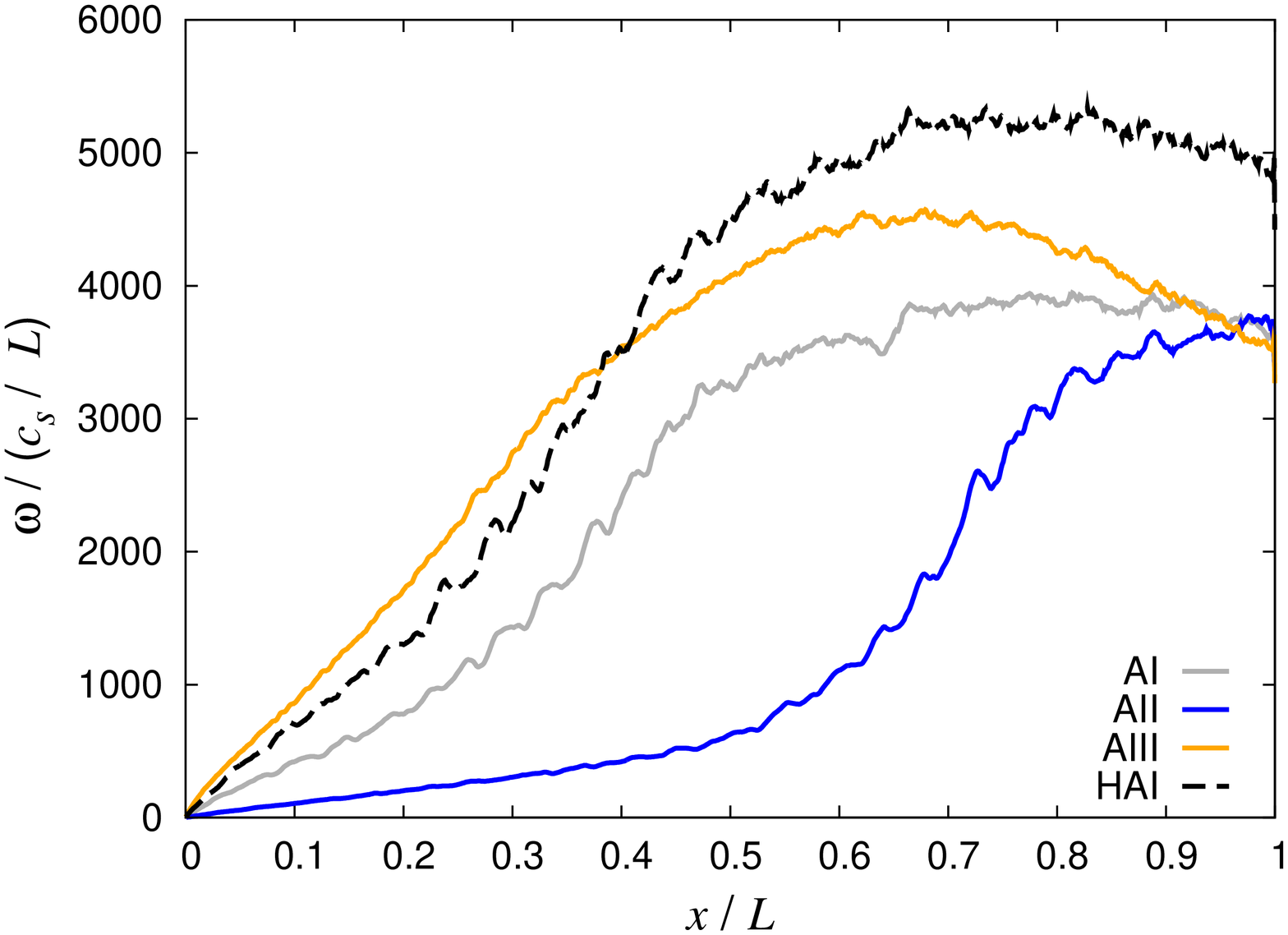} \\
\includegraphics[width=.7\linewidth, trim= 0cm 0cm 0cm 0cm, clip=true,angle=270]{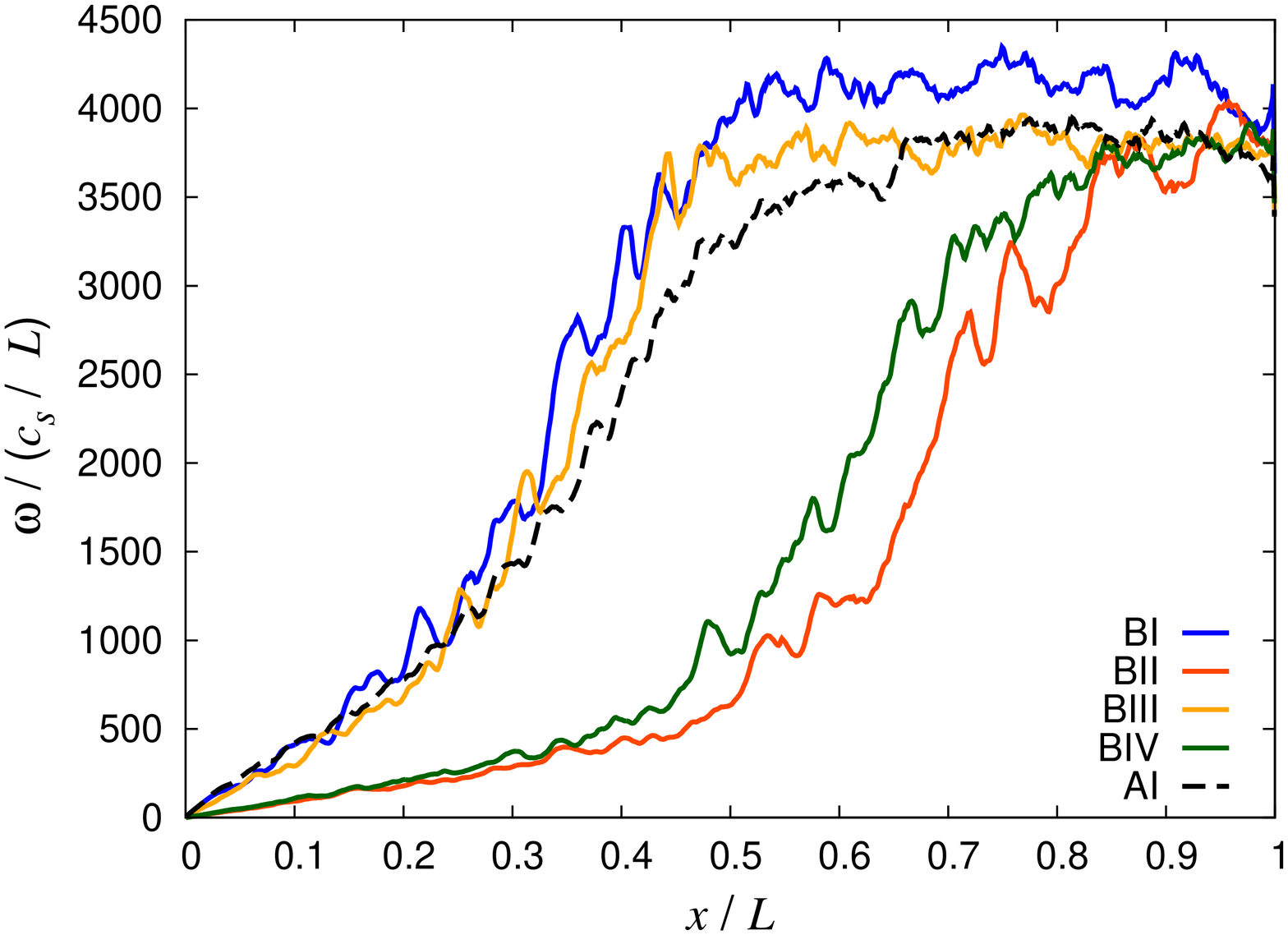} 
\caption{Profile of the $rms$ vorticity (averaged in the $yz$-plane) for models A ({\it top}) 
and models B ({\it bottom}). The profiles were averaged in time from $t=3L/v_0$ to $t=4L/v_0$.}
\label{fig:omg-a1}
\end{figure}

Figure~\ref{fig:omg-a1} shows the profile of the vorticity intensity $\omega$ for the models in Table 1. The comparison between models AI and AII (top panel) shows that vorticity is generated at a faster rate with distance (along the $x$-direction) for models with the higher values of $\delta \rho_{rms}/\rho_0$. Comparing models with the same density contrast but different density structures size (models AI and AIII),
the vorticity increases with distance more quickly for the model with smaller
$\lambda_{\rho}$. We also observe that the vorticity intensity $\omega$, after a phase of fast increase in the beginning of the box, saturates (or increases very slowly) for the models with the highest values of $\delta \rho_{rms}/\rho_0$. The models with simulated turbulence (bottom panel) show the same behaviour as models A (model AI is shown in dashed line for comparison), with the rate of vorticity generation being higher for the models with higher density contrast (models BI and BIII).

\begin{figure}
\centering
\includegraphics[width=.7\linewidth, trim= 0cm 0cm 0cm 0cm, clip=true,angle=270]{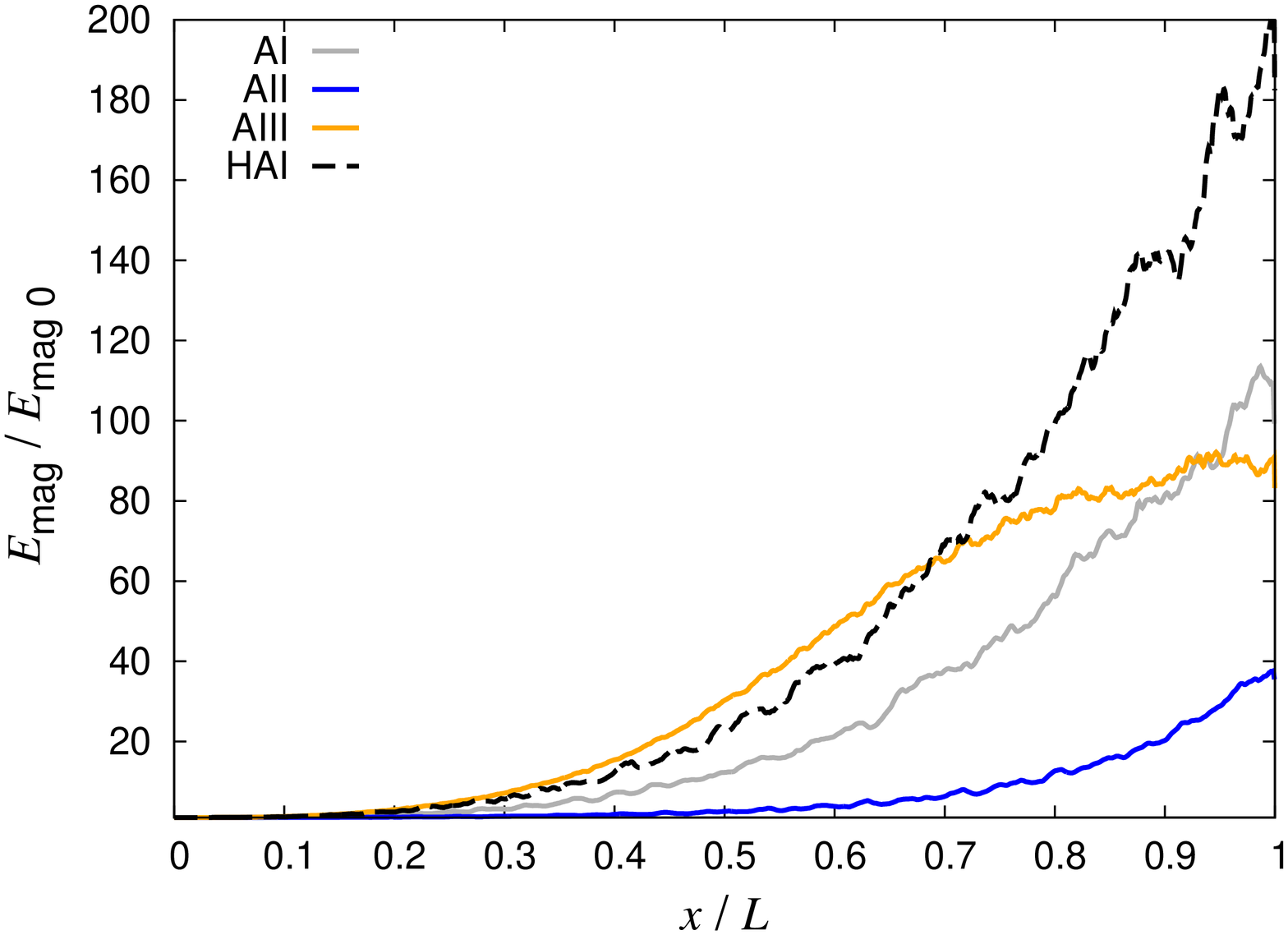} \\
\includegraphics[width=.7\linewidth, trim= 0cm 0cm 0cm 0cm, clip=true,angle=270]{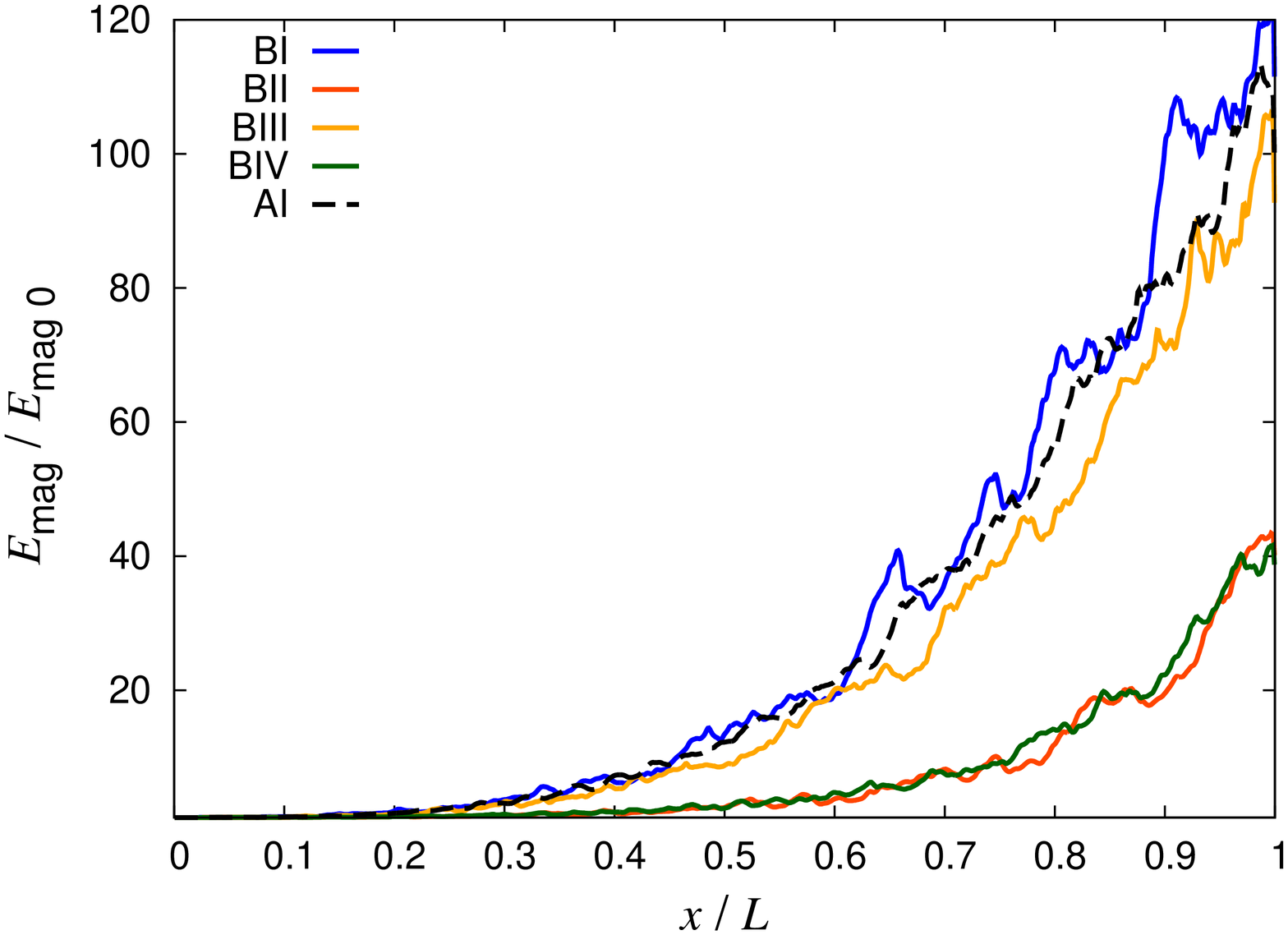} 
\caption{Profile of the magnetic energy density for models A ({\it top}) 
and models B ({\it bottom}). The profiles were averaged in time from $t=3L/v_0$ to $t=4L/v_0$.}
\label{fig:rho-rms}
\end{figure}

The magnetic energy profiles are shown in Figure~\ref{fig:rho-rms}. We can infer for  models A (top panel) that the rate of magnetic energy growth per distance is larger for models AI and AIII; models with different density structure sizes also show differences in the magnetic field evolution: it is faster for the models with smaller $\lambda_{\rho}$ at the beginning of the box, but the magnetic energy achieves a maximum and then saturates before the end of the box is reached. The final magnetic energy (at $x=L$) differs greatly between models AI and AII, but it is similar between  models with different values of $\lambda_{\rho}$ (AI and AIII). Models AI and AII differ in $\delta \rho_{rms}/\rho_0$ by a factor of $3$, and the final magnetic energy density increases by a factor of $\approx 115$ in the first case and $\approx 39$ in the second one. The same dependence of the magnetic energy growth rate on $\delta \rho_{rms}/\rho_0$ is observed in models B (Figure~\ref{fig:rho-rms}, bottom panel).

\begin{figure}
\centering
\includegraphics[width=.7\linewidth, trim= 0cm 0cm 0cm 0cm, clip=true,angle=270]{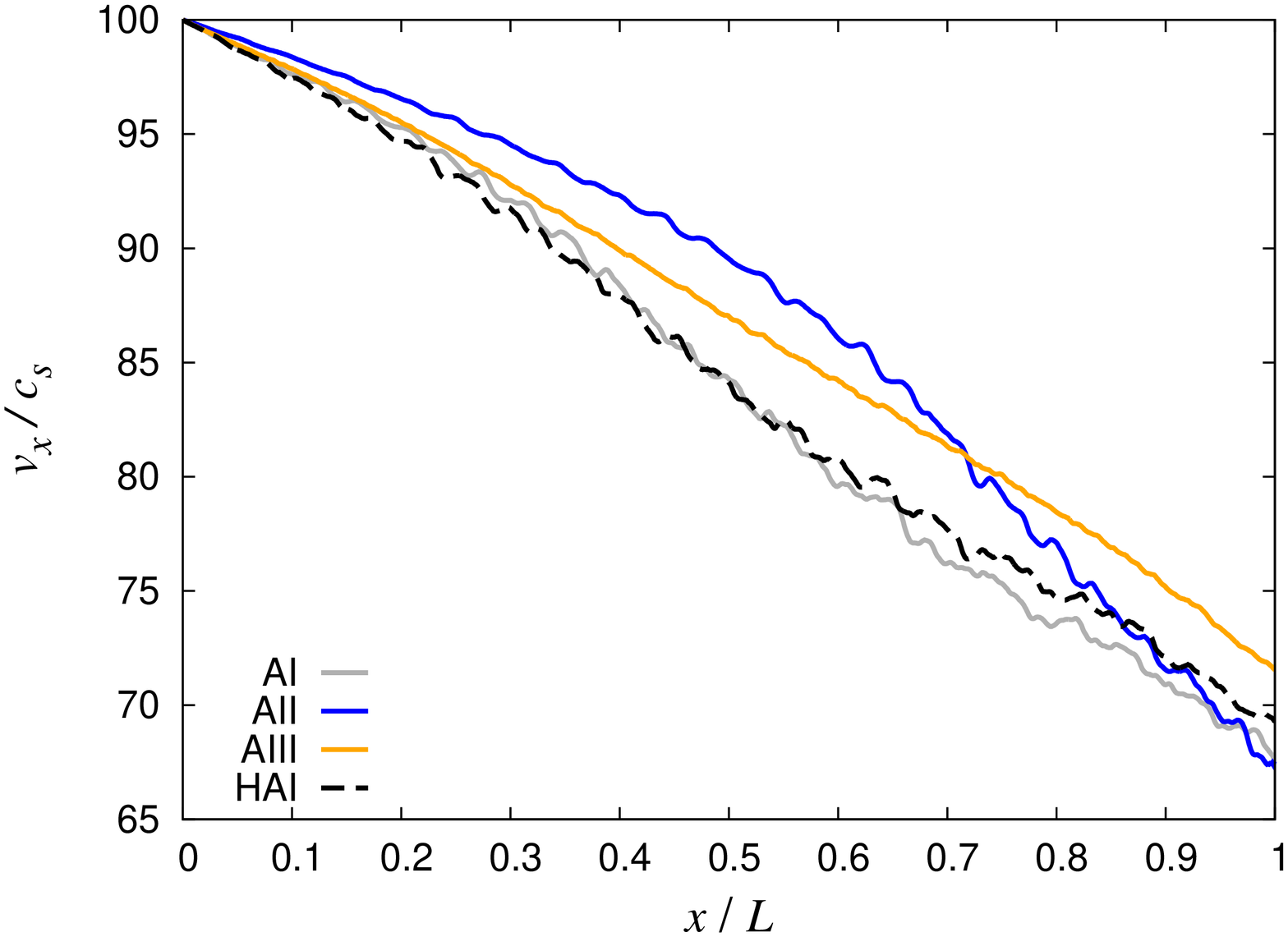} \\
\includegraphics[width=.7\linewidth, trim= 0cm 0cm 0cm 0cm, clip=true,angle=270]{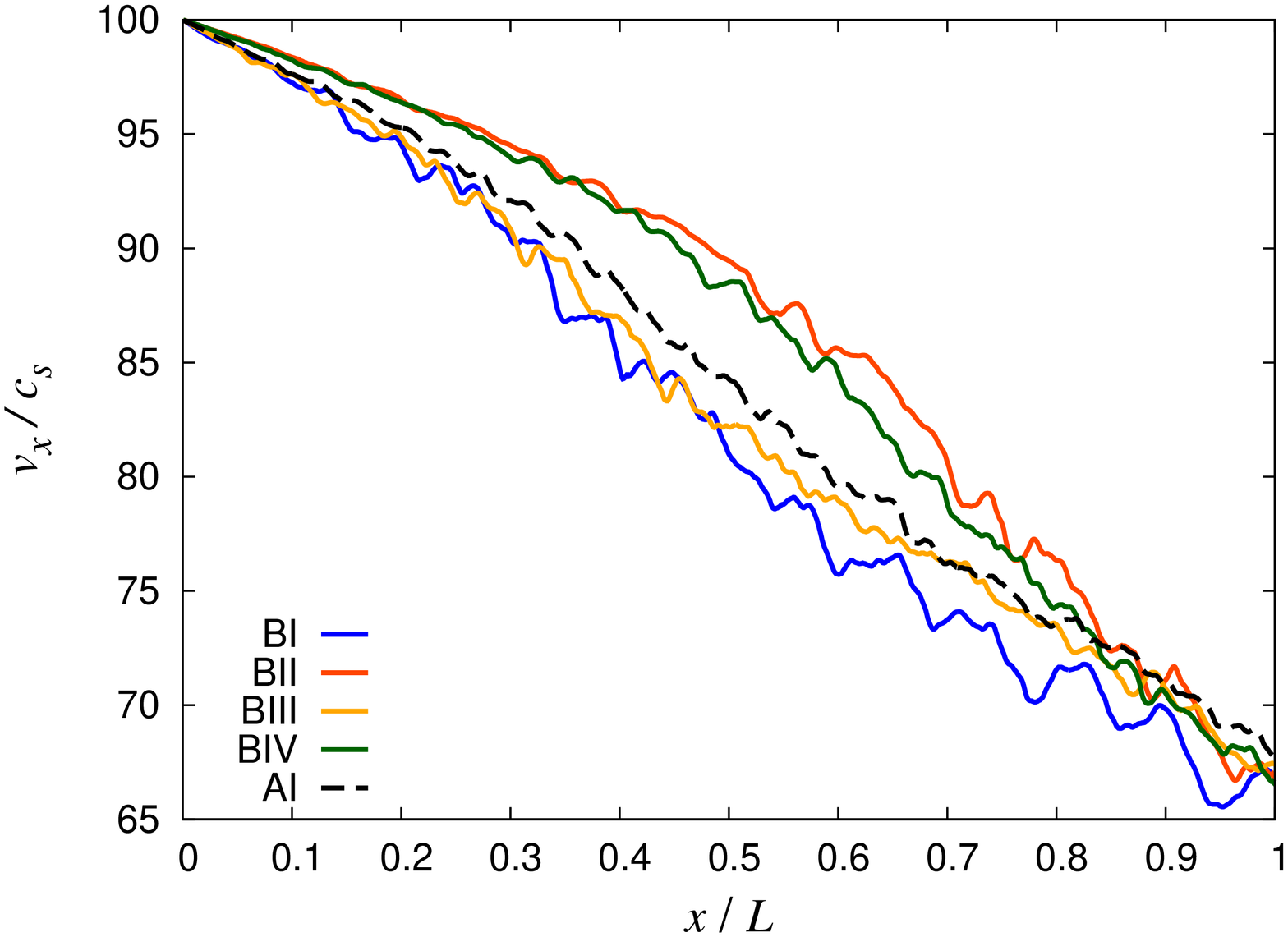}
\caption{Profile of the $x$-component of the velocity for models A ({\it top}) 
and models B ({\it bottom}). The profiles were averaged in time from $t=3L/v_0$ to $t=4L/v_0$.}
\label{fig:velxA}
\end{figure}

Part of the magnetic field amplification observed in Figure~\ref{fig:rho-rms} 
is due to plasma compression. As the CR pressure slows down the incoming flow in the precursor  the plasma is compressed by a certain factor, as a consequence the magnetic field (which is initially perpendicular to the shock) increases its intensity by approximately this same factor. Figure~\ref{fig:velxA} shows the 
profiles of the mean velocity for models A (top panel) and B (bottom panel).

The initial velocity is reduced by $\approx 30\%$ of its initial value in all models; 
at the same time that density is increased by a similar factor (not shown). 
Therefore, the compression can  account only for an increase of a factor $\approx 1.7$ in the magnetic energy density, and the remaining being due to the small-scale dynamo.

A key ingredient in the BJL09 theory is the generation of solenoidal motions within the precursor, which is the turbulent velocity component responsible for amplifying the magnetic field. In order to separate the solenoidal component $\mathbf{v}_{\rm s}$ ($\mathbf{\nabla \cdot v_s} = 0$) of the velocity field from its curl-free component $\mathbf{v}_{\rm p}$ ($\mathbf{\nabla \times v_p} = 0$), we calculate the potential fields $\phi$ and $\mathbf{A}$ from the the Helmholtz decomposition:
\begin{equation}
\mathbf{v} = \nabla \phi + \nabla \times \mathbf{A}, 
\end{equation}
with $\nabla \phi = \mathbf{v}_{\rm p}$ and $\nabla \times \mathbf{A} = \mathbf{v}_{\rm s}$.

\begin{figure*}
\begin{tabular}{c c}
\includegraphics[width=.35\linewidth, trim= 0cm 0cm 0cm 0cm, clip=true,angle=270]{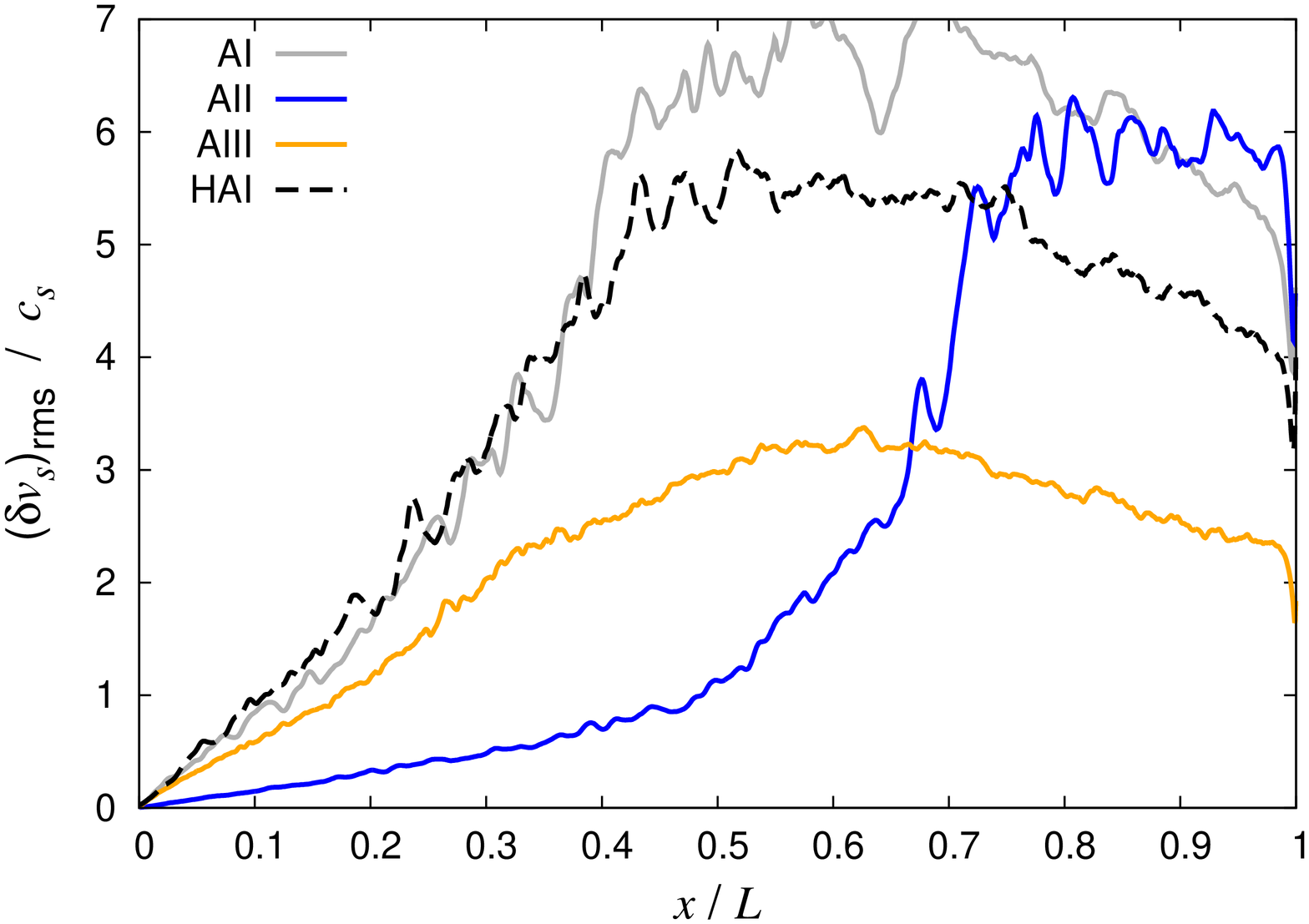} &
\includegraphics[width=.35\linewidth, trim= 0cm 0cm 0cm 0cm, clip=true,angle=270]{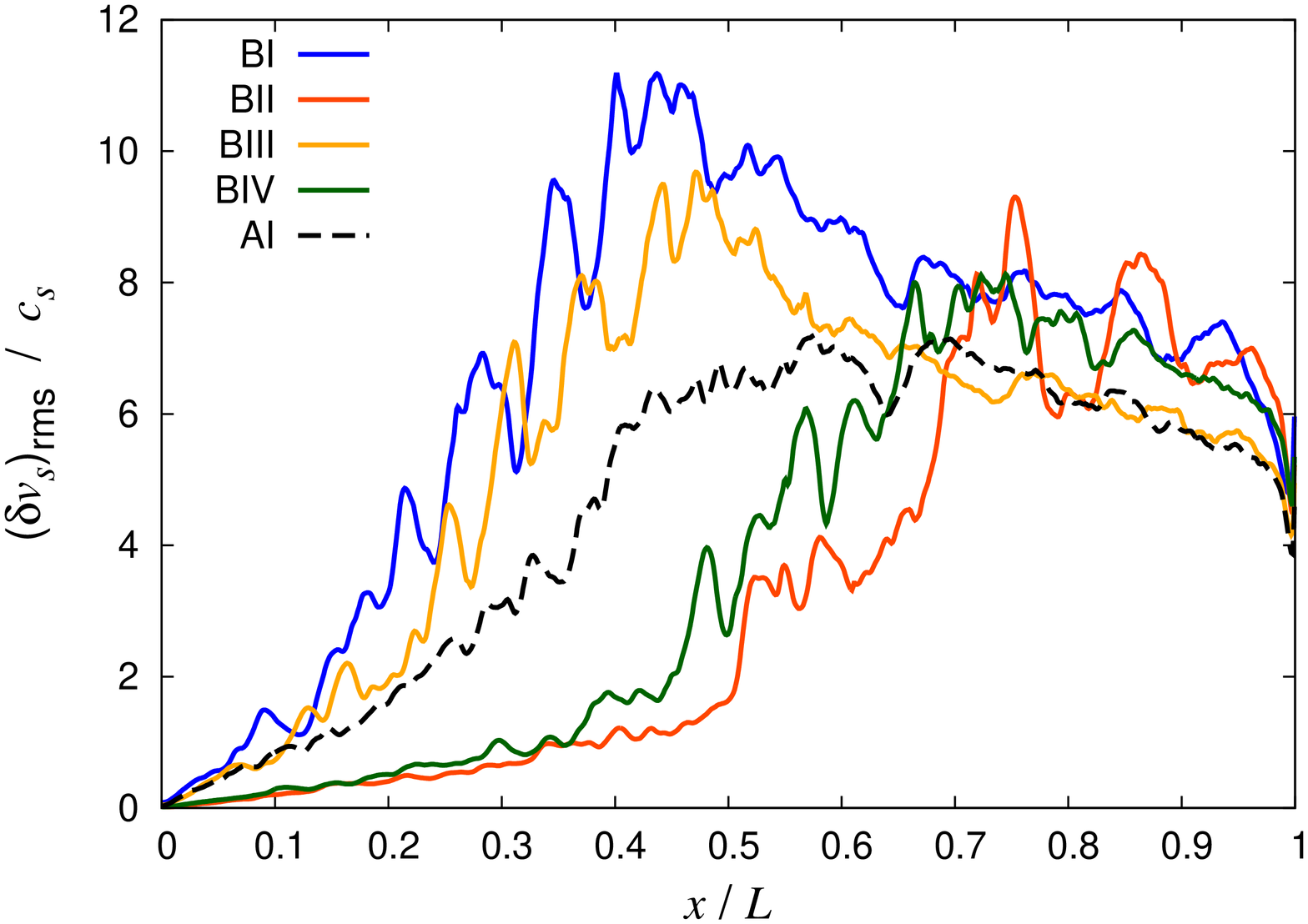} \\
\includegraphics[width=.35\linewidth, trim= 0cm 0cm 0cm 0cm, clip=true,angle=270]{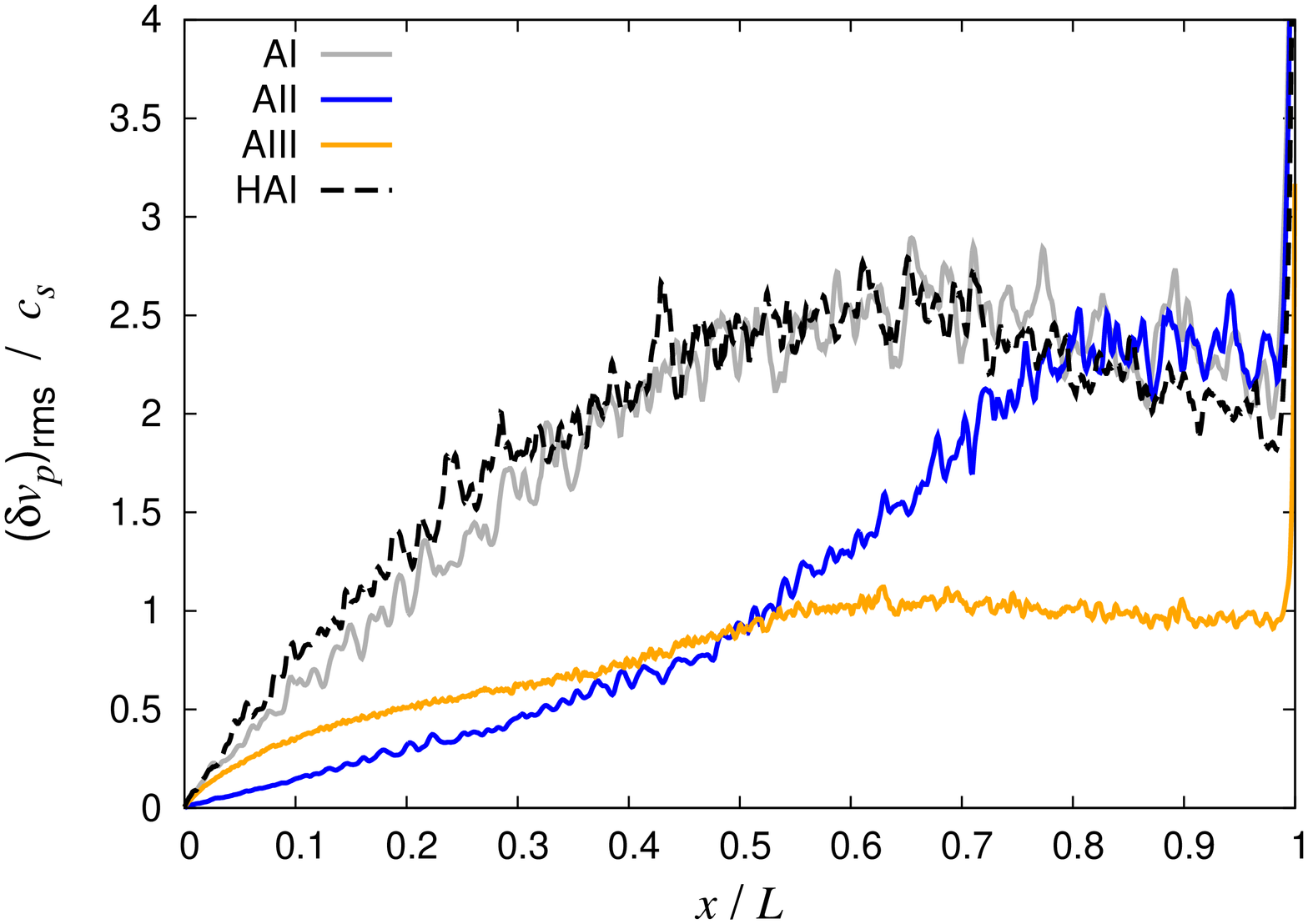} &
\includegraphics[width=.35\linewidth, trim= 0cm 0cm 0cm 0cm, clip=true,angle=270]{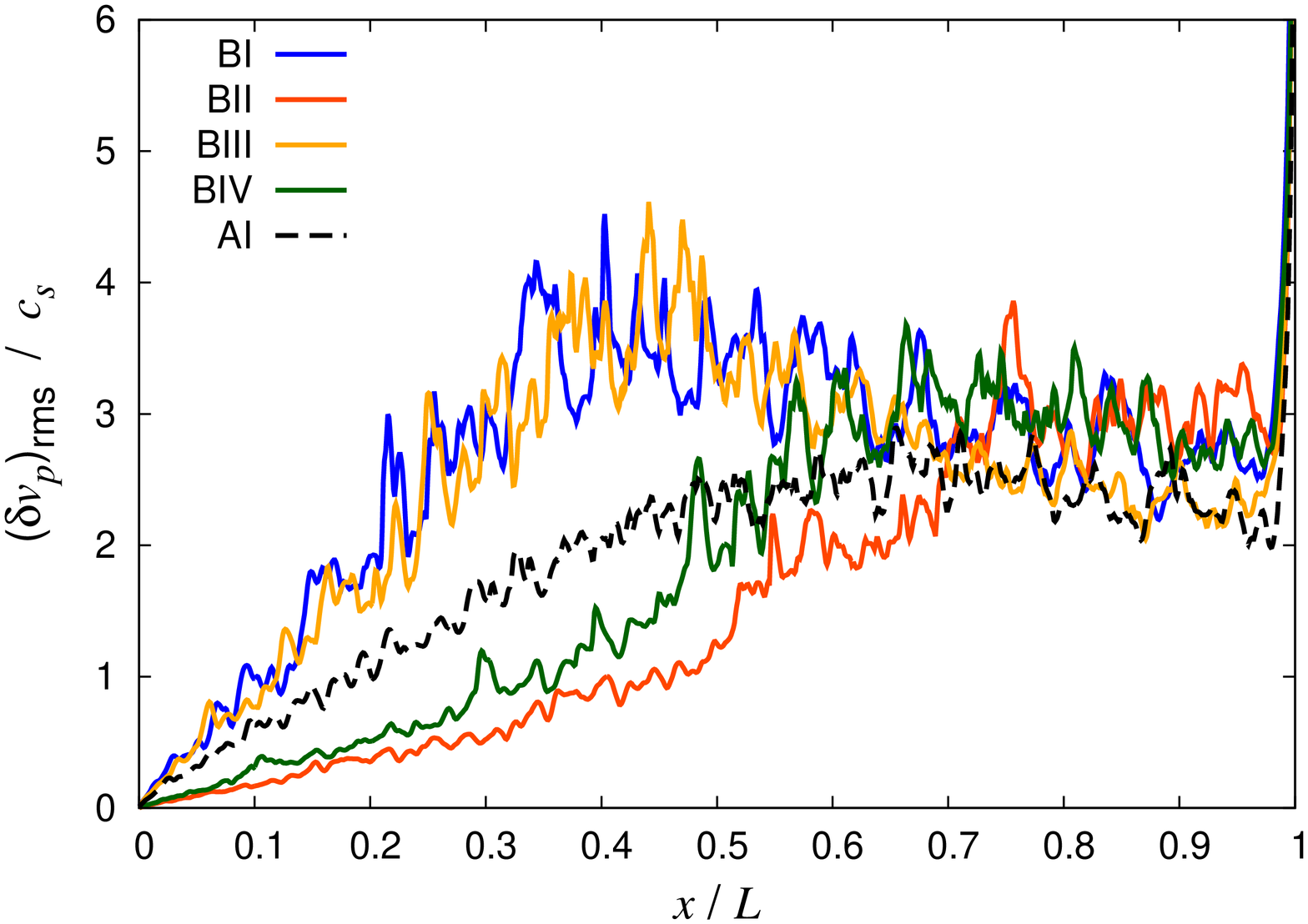} \\
\includegraphics[width=.35\linewidth, trim= 0cm 0cm 0cm 0cm, clip=true,angle=270]{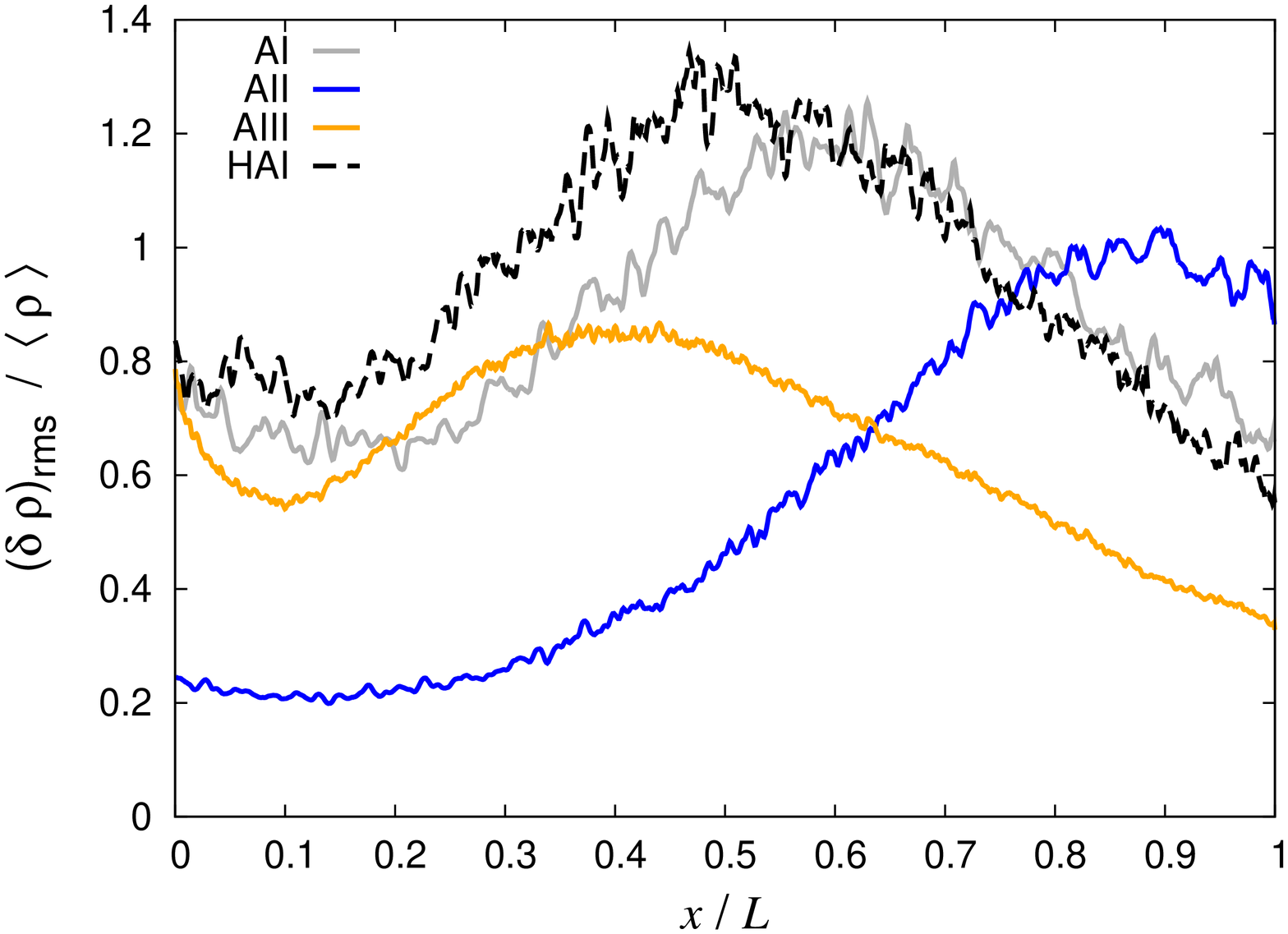} &
\includegraphics[width=.35\linewidth, trim= 0cm 0cm 0cm 0cm, clip=true,angle=270]{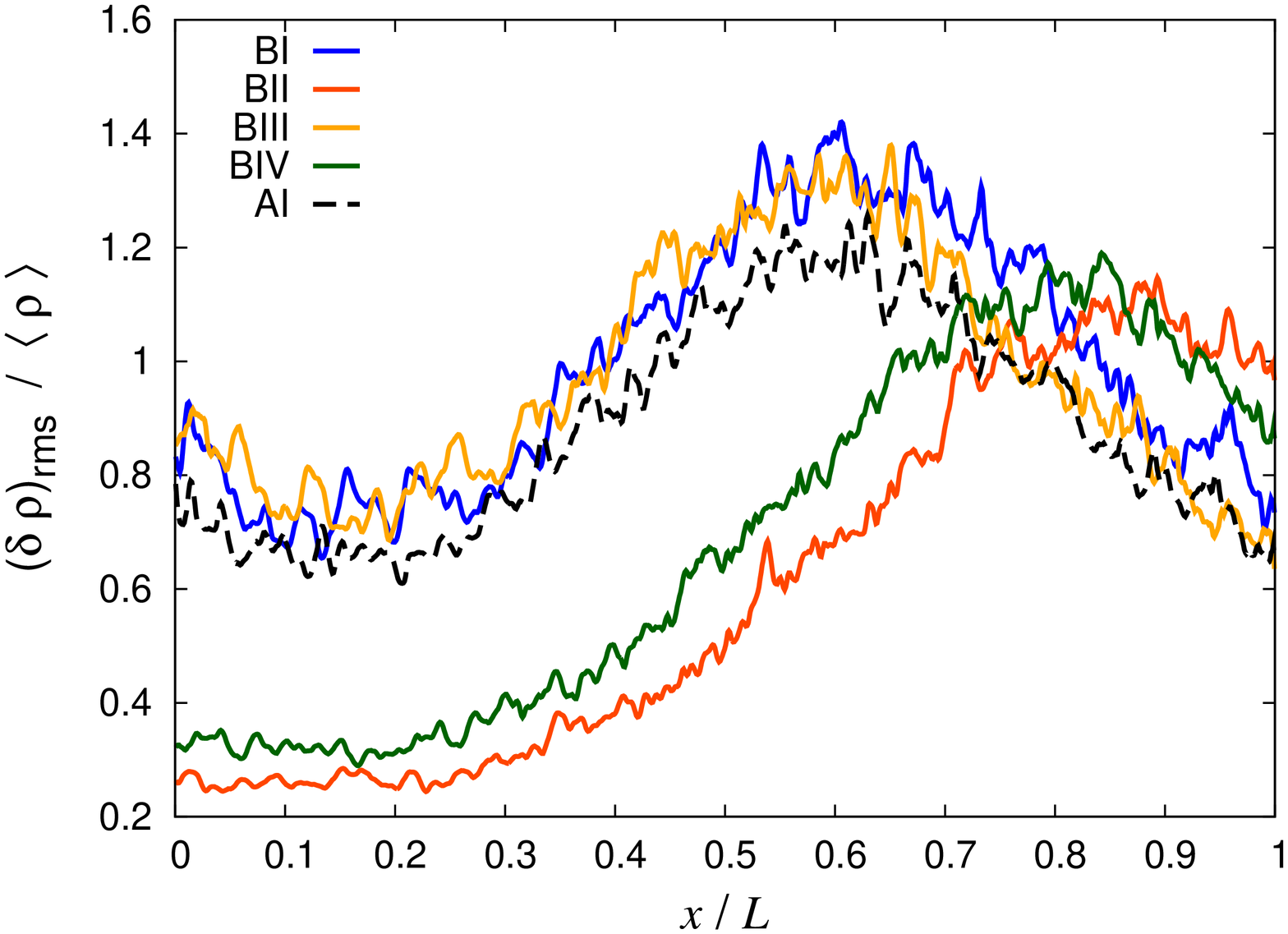}
\end{tabular}
\caption{Profile of the $rms$ solenoidal velocity fluctuations $\delta \mathbf{v}_s$ (top row), 
the $rms$ potential velocity fluctuations $\delta \mathbf{v}_p$ (middle row), 
and density fluctuations $\langle \delta \rho \rangle / \langle \rho \rangle$ (bottom row).
Left column: models A; right column: models B. 
The profiles were averaged in time from $t=3L/v_0$ to $t=4L/v_0$.}
\label{fig:decos}
\end{figure*}

Top row of Figure~\ref{fig:decos} shows the profile of the $rms$ of the velocity fluctuations $\delta \mathbf{v}_s$ for models A (left) and models B (right). Comparing models AI and AII, we clearly see that the solenoidal velocity fluctuation is higher for model AI, with  higher $\delta \rho_{rms}/\rho_0$. In the middle of the box ($x/L = 0.5$), the solenoidal fluctuations for these models are  $\approx 1 \; c_s$ (AII) and $6\; c_s$ (AI), although at the end of the box ($x=L$) the velocities are approximately equal ($ \approx 6\; c_s$). The solenoidal velocity of model AII initially increases more slowly and then increases at a higher rate (after $x/L=0.5$). The model AIII, with smaller $\lambda_{\rho}/D_{CR}$, on the other hand, presents slower growth of  velocity fluctuations after the middle of the box is achieved. The models B (right column) show a trend similar to that of models A in the growths of solenoidal velocity fluctuations: an initially slow growth followed by a faster growth for models with smaller $\delta \rho_{rms}/\rho_0$ (models BII and BIV). The models BI and BIII show a fast increase until $x \approx 0.5-0.6 L$, where the maximum is achieved; after that a fast decrease by a factor of approximately 2 is followed, to finally decrease slowly at the end of the box.

In the middle row of Figure~\ref{fig:decos}, the profiles of the $rms$ of the potential velocity fluctuations $\delta \mathbf{v}_p$ are shown. The potential fluctuations achieve nearly the same saturation level $\delta v_{p, rms} \approx 2\; c_s$ for  models AI and AII. For model AII, with smaller $\delta \rho_{rms}/\rho_0$, the saturation level is only achieved at larger values of $x$ ($x \approx 0.7$). The model AIII with smaller $\lambda_{\rho}$, on the other hand, has lower maximum values of the potential fluctuations, compared to the reference model AI ($\approx 2.5$ for model AI, and $\approx 1$ for model AIII). Once more  models B (right column) follow similar trends as models A, with the density contrast ruling the behaviour of the models.

The potential velocity fluctuations are important to generate more density structures along the preshock, otherwise the solenoidal turbulent motions will operate only in a small fraction of the preshock, because besides amplifying the magnetic field, the solenoidal turbulent motions act to homogenize the density field (due to turbulent mixing). The last row of Figure~\ref{fig:decos} shows the profiles of the $rms$ of the density fluctuations $\delta \rho$. Comparing models A (left column), it can be seen that after an interval in which the density fluctuations remain constant (until $x/L \approx 0.3$ for model AII), they grow at nearly the same rate for models AI and AII, and after achieving a maximum value, the fluctuations start to decrease. The maximum values are higher and are achieved first for models with higher initial density contrast. In the case of model AIII, after the maximum value is achieved in the density contrast at $x/L \approx 0.5$, it  diminishes. This maximum value is smaller as smaller the $\lambda_{\rho}$, and it should be remarked that the values of the density fluctuations at the end of the box are smaller than the initial values, meaning that density fluctuations are not efficiently generated by the induced turbulence, in agreement with the potential velocity fluctuations (middle row of Figure~\ref{fig:decos}). Figures~\ref{map-final-rho} and~\ref{map-final-emag} show the density and magnetic energy density distribution, respectively, in the central  $xy$-plane for models AI and BI, at the time $t = 4 L/v_0$.

\subsection{Power spectrum}

The turbulent dynamo, usually studied in the incompressible limit of the MHD equations, amplifies the magnetic field energy from the smaller scales to the larger scales (see \cite{2015ASSL..407..163B} for a recent review). The saturation is expected to happen scale-by-scale (first at the smaller scales), when the magnetic energy achieves near equipartition with the kinetic energy. For the incompressible reference case, it is observed that in the saturation of the dynamo, the power spectrum of the magnetic field (magnetic energy) is close to the velocity field power spectrum (kinetic energy).

\begin{figure}
\begin{center}
\begin{tabular}{c c}
\includegraphics[width=.7\linewidth, trim= 0cm 0cm 0cm 0cm, clip=true,angle=270]{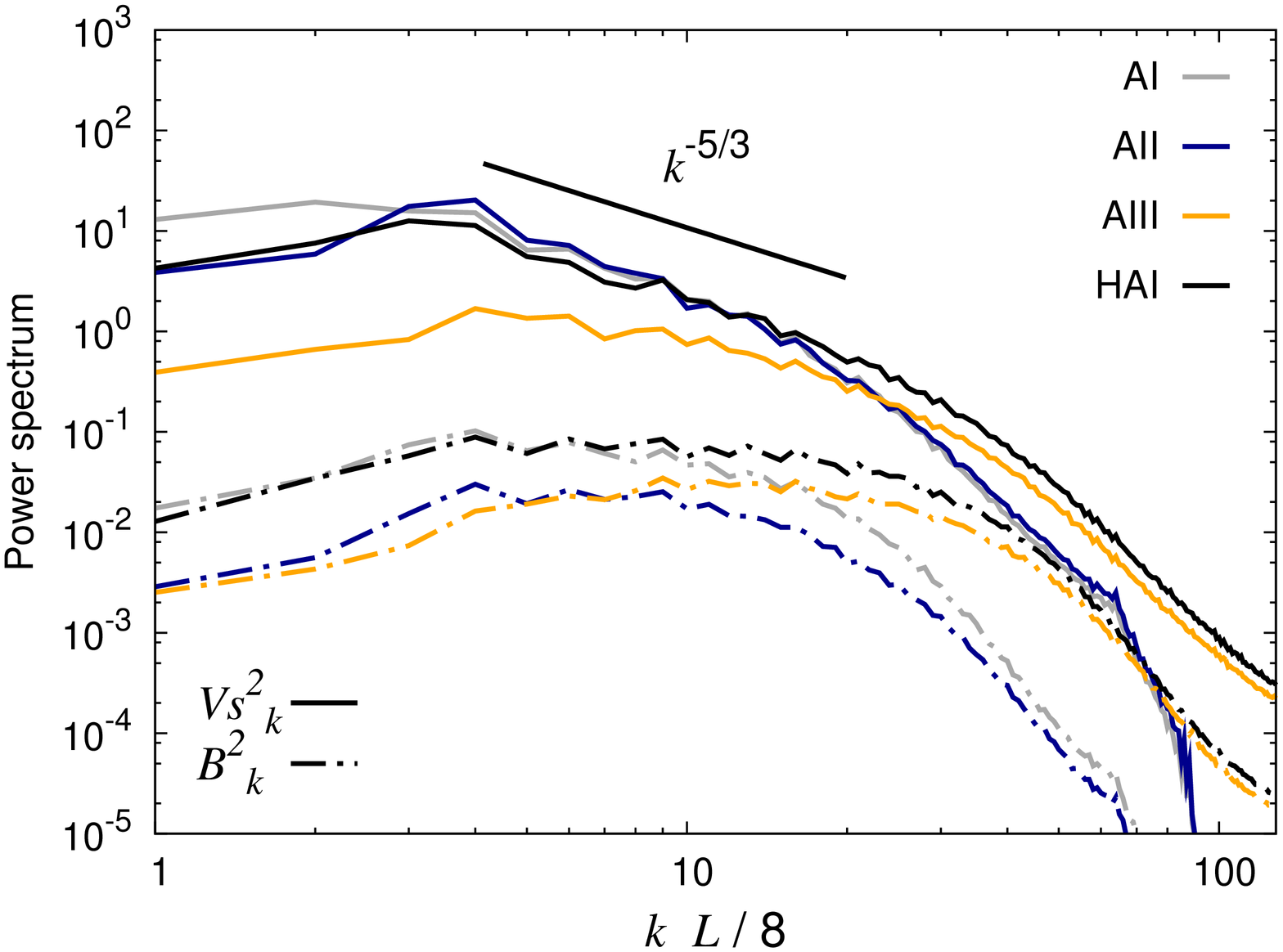} \\
\includegraphics[width=.7\linewidth, trim= 0cm 0cm 0cm 0cm, clip=true,angle=270]{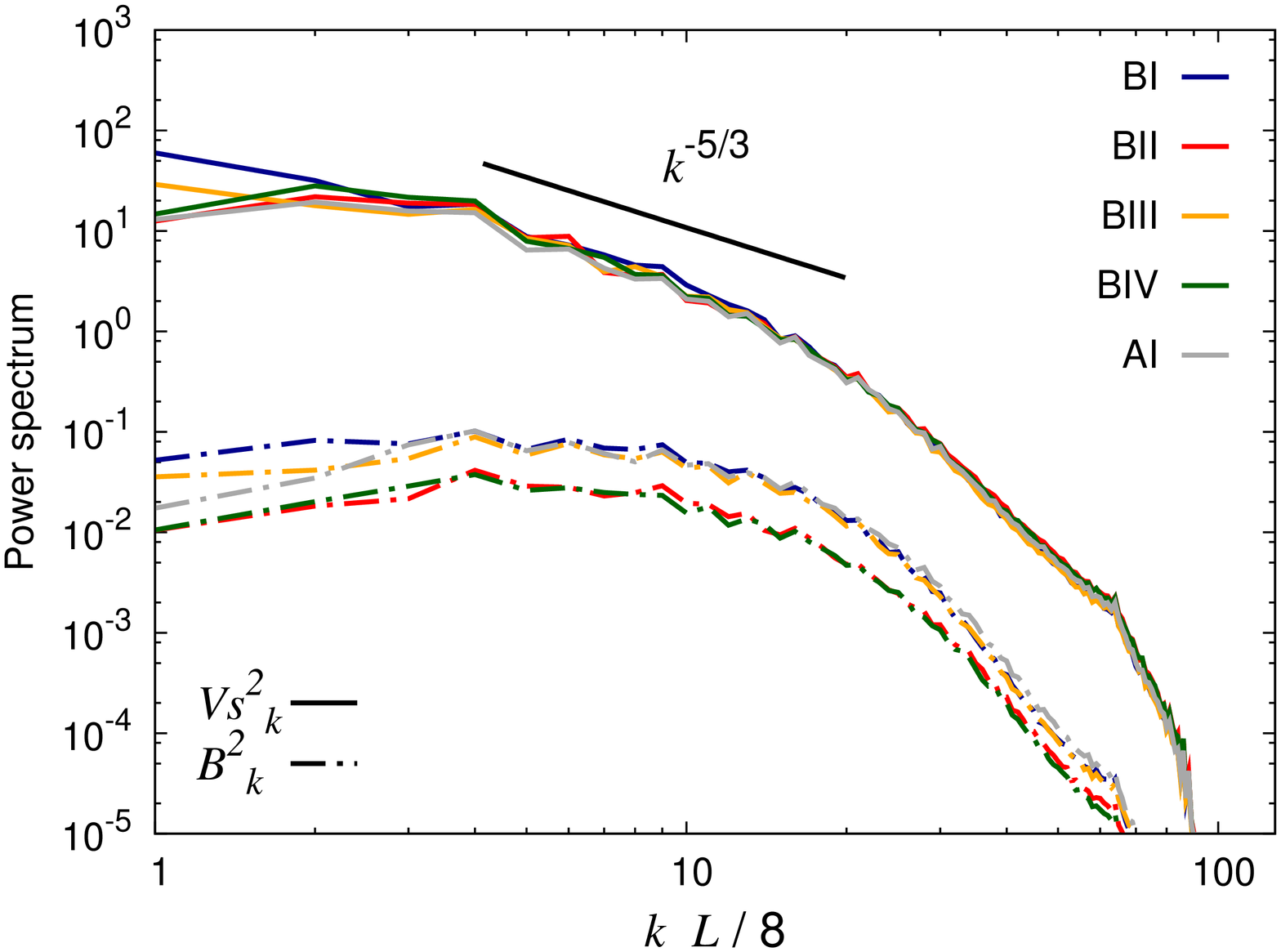}
\end{tabular}
\end{center}
\caption{Power spectrum of the magnetic field $\mathbf{B}$ (dashed lines) and of the solenoidal 
velocity field $\mathbf{v}_s$ (continuous line) for models A ({\it top}) and models B ({\it bottom}). 
The power spectrum were calculated in the $yz$-plane of the last grid position in the $x$-direction.
The power spectra were averaged in time from $t=3L/v_0$ to $t=4L/v_0$.}
\label{fig-ps1}
\end{figure}

Figure~\ref{fig-ps1} shows a comparison between the power spectrum of the magnetic field $\mathbf{B}$ with the power spectrum of the solenoidal velocity field $\mathbf{v}_s$, for models A (top panel) and models B (bottom panel).  The power spectra are calculated in the $yz$-plane at the last cell in the $x$-direction of the domain. The Kolmogorov power law $\propto k^{-5/3}$ is also shown for comparison. We observe that for all A models  the solenoidal velocity field power spectrum peaks approximately at the wave-number corresponding to the maximum of the initial density  structures ($k_{\max} \approx L_y/\lambda_{\rho}$). At these scales, 
the magnetic power spectrum for all models A is approximately 2 orders of magnitude lower than the solenoidal velocity power spectrum. Model AII presents  magnetic field power spectrum  smaller than those of model AI. Model AIII, when compared to model AI, presents the power spectrum of the magnetic field closer to the power spectrum of the solenoidal velocity component, that is, closer to the equipartition at the smaller scales. Model B (bottom panel) presents the power spectrum for the magnetic field lower for models with lower density contrast. However, contrary to models A, their power spectra present more power at 
the larger scales (smaller $k$), especially for models BI and BIII. The difference 
between the magnetic and velocity power spectra is about 1 order of magnitude for larger values of $k$ ($k>20$); it is approximately 2 orders of magnitude for smaller values of the wave-number ($5>k>10$), and the maximum difference of about 3 orders of magnitude is reached for $k=1$.

\begin{figure*}
\begin{tabular}{c c}
\includegraphics[width=.35\linewidth, trim= 0cm 0cm 0cm 0cm, clip=true,angle=270]{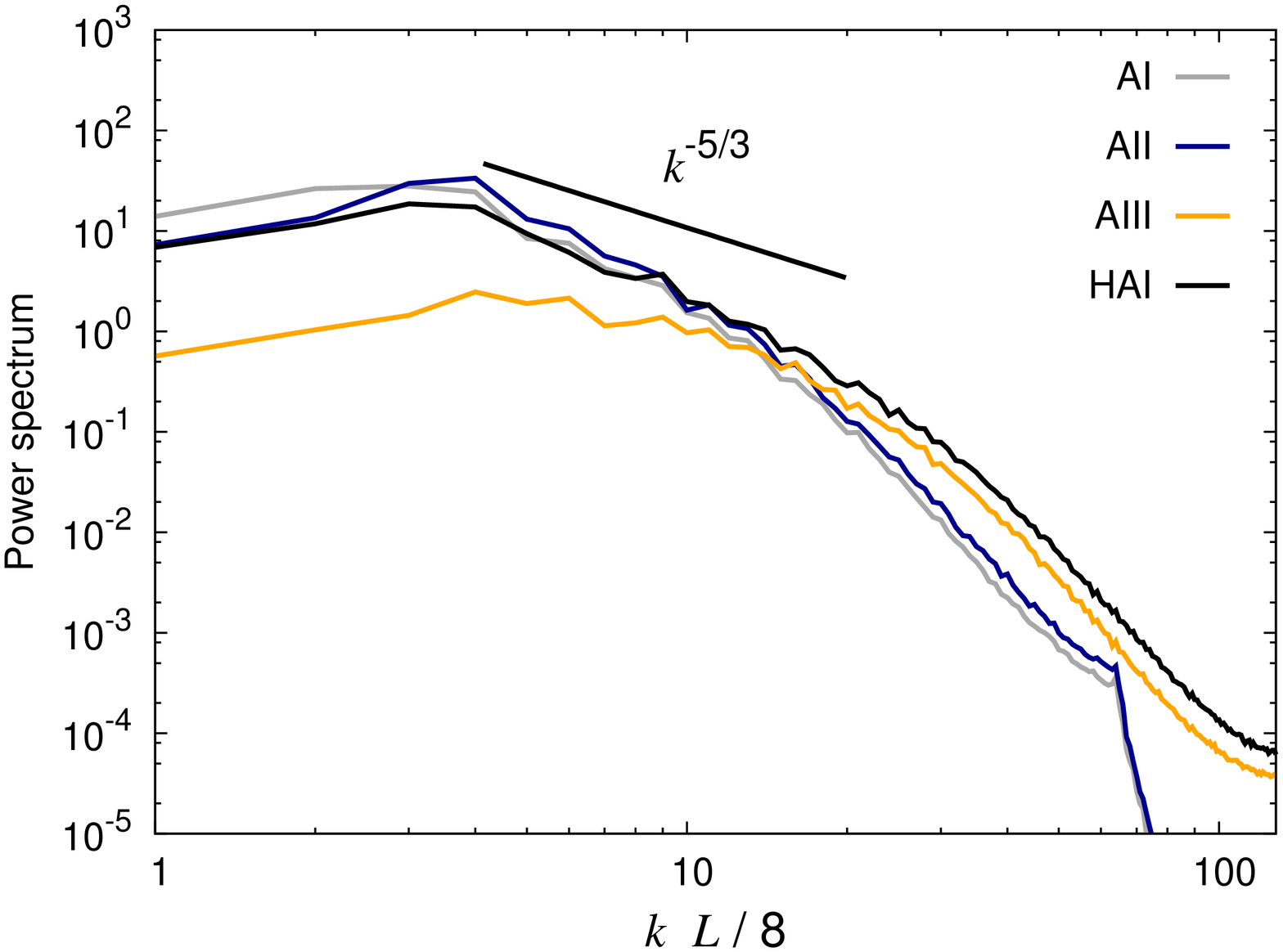} &
\includegraphics[width=.35\linewidth, trim= 0cm 0cm 0cm 0cm, clip=true,angle=270]{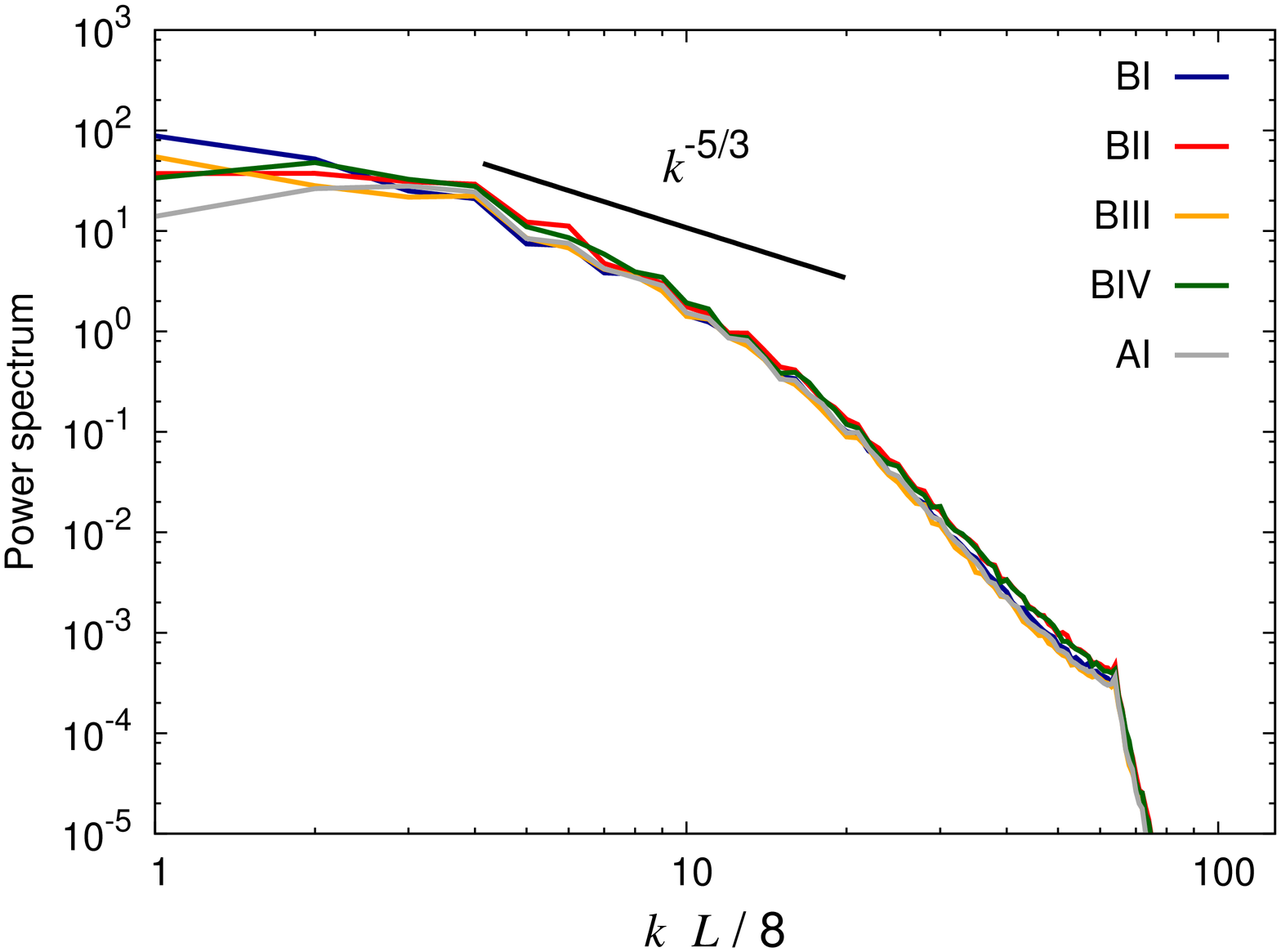} \\
\includegraphics[width=.35\linewidth, trim= 0cm 0cm 0cm 0cm, clip=true,angle=270]{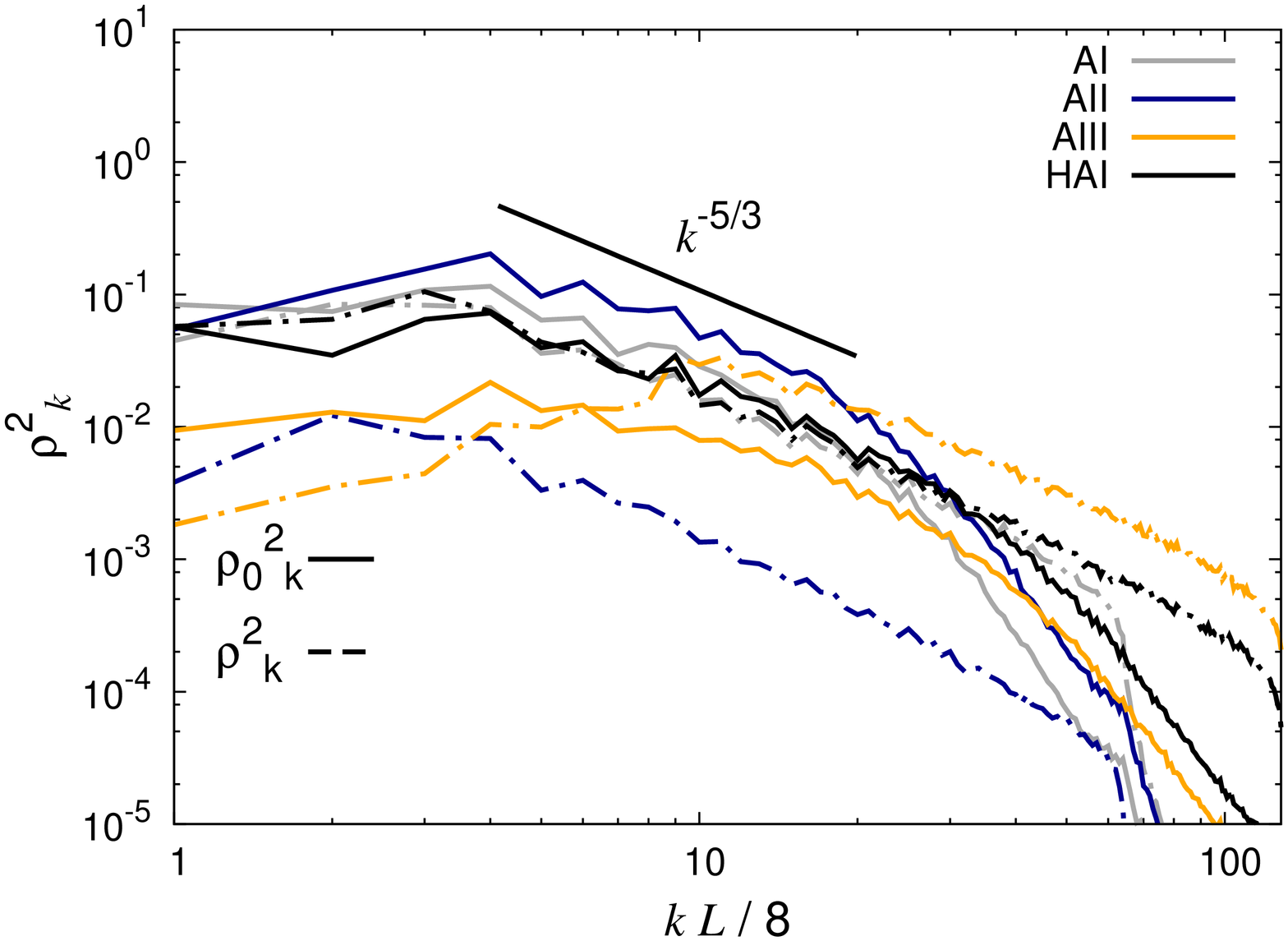} &
\includegraphics[width=.35\linewidth, trim= 0cm 0cm 0cm 0cm, clip=true,angle=270]{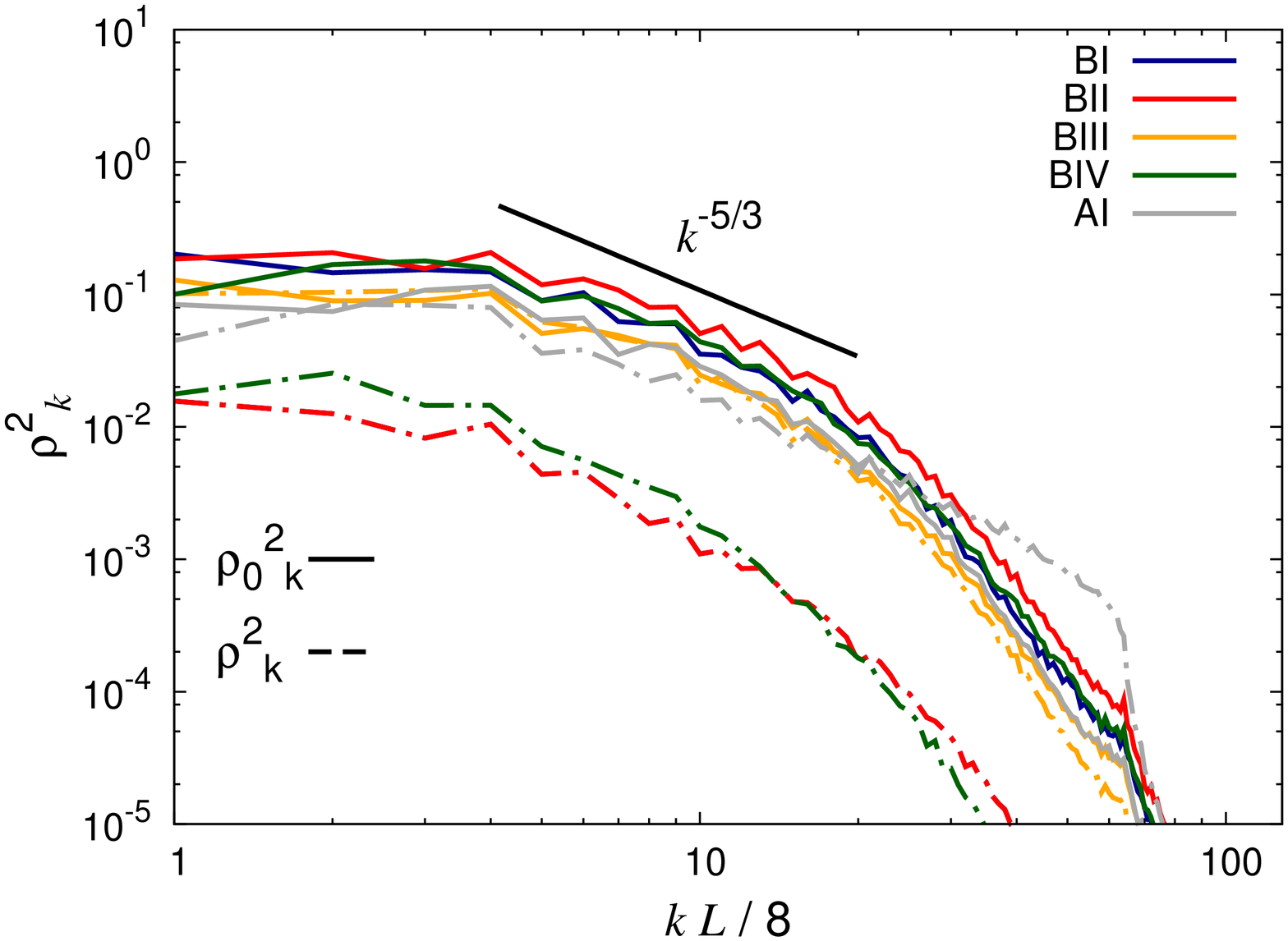}
\end{tabular}
\caption{Power spectrum of the potential velocity field $\mathbf{v}_p$ (top row) and of the 
density field $\rho$ (bottom row) for models A ({\it left column}) and models B ({\it right column}). 
The power spectrum were calculated in the $yz$-plane of the last grid position in the $x$-direction.
The density power spectra of the first grid position is shown for comparison.
The power spectrum were averaged in time from $t=3L/v_0$ to $t=4L/v_0$.}
\label{fig:ps_vp}
\end{figure*}

Figure~\ref{fig:ps_vp} shows the power spectrum of the potential velocity field (top row) and of the density (bottom row). Comparing the power spectrum of the potential velocity field of models AI and AII, can be seen that they are nearly identical, peaking at the same wavenumber as the solenoidal velocity field does ($k \approx 4$). Model AIII differs from  model AI in the power which is smaller for 
$k<10$,  and nearly constant. In the case of models B (left column), for $k>2$, they have similar potential velocity spectra. 

Comparing models AI and AII we see that the density power spectrum (left column in Figure~\ref{fig:ps_vp}) close to the shock are similar; the power increases  at the larger scales for the model  with smaller initial $\delta \rho_{rms}/\rho_0$ (AII). The peak in the final power spectra seems to be localized close to the wave-number where the initial density spectra peaks $k_{\max} \approx L_y/\lambda_{\rho} = 2.5$. We also observe that the power at the smaller scales decreases. This is probably attributed to the numerical dissipation, which dumps the velocity fluctuations and consequently the density fluctuates over these scales. Now comparing models models AI and AIII, we do see the final spectrum of model AIII to increase for scales larger than $\lambda_{\rho}$ ($k < k_{\min}$). All models B present a very similar final density power spectra. 

%%==============================================================================
%%
%%
\section{Discussion}

\label{sec:discussion}

\begin{figure}
\begin{center}
\begin{tabular}{c c}
\includegraphics[width=.7\linewidth, trim= 0cm 0cm 0cm 0cm, clip=true,angle=270]{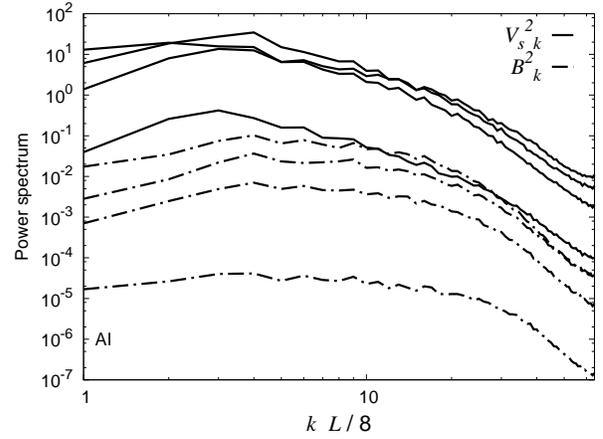}
\end{tabular}
\end{center}
\caption{Power spectrum of the magnetic field $\mathbf{B}$ (dashed lines) and of the solenoidal 
velocity field $\mathbf{v}_s$ (continuous line) for model AI. 
Each curve represents the power spectrum calculated in the $yz$-plane of a different position in the 
$x$-direction. The lowermost curves are for position $x/L=0.1$, and the uppermost correspond to $x/L \approx 1$.
The intermediary curves are separated in the $x$ coordinate by $\Delta x/L = 0.3$.
The power spectra were averaged in time from $t=3L/v_0$ to $t=4L/v_0$.}
\label{fig:ai-ps_evol}
\end{figure}

The comparison between the power spectrum of the solenoidal velocity and of the magnetic field shows that, for all the parameters considered in this work, the resulting amplification of the magnetic field close to the shock is far from equipartition with the kinetic energy available from the solenoidal velocity field, even for the smaller scales solved in our simulations (Figure~\ref{fig-ps1}). The evolution of each power spectrum (solenoidal velocity and magnetic field) along the distance inside the precursor is shown for model AI in Figure~\ref{fig:ai-ps_evol}. It can be seen that the velocity power is increasing inside the precursor at approximately the same rate for all the wavenumbers.

The role of the potential (compressible) component of the velocity field in the turbulent amplification of the magnetic field is not clear, as the dynamo studies usually consider the incompressible limit of the MHD equations. Despite of this, the solenoidal velocity fluctuations dominate over the compressible ones in the turbulence generated (Figure~\ref{fig:decos}).

Additionally, the total velocity fluctuations achieves only a few percent of the upstream flow velocity $v_0$, and its evolution is shown here to depend strongly on the initial density contrast, as well as on the density structures size (Figure~\ref{fig:decos}). The saturation values are in the range $0.04-0.1 v_0$ for the models we present, and are higher for the highest values of the initial density contrast and of the density structures size. 

\begin{figure}
\begin{center}
\begin{tabular}{c c}
\includegraphics[width=.7\linewidth, trim= 0cm 0cm 0cm 0cm, clip=true,angle=270]{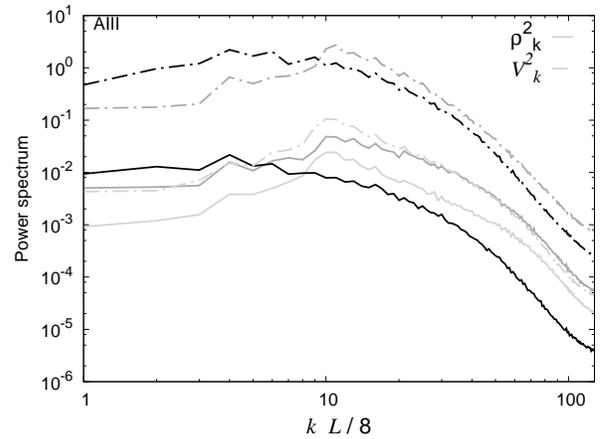}
\end{tabular}
\end{center}
\caption{Power spectrum of the density field $\rho$ (dashed lines) and of the  
velocity field $\mathbf{v}$ (continuous line) for model AIII. Each curve represents the power spectrum calculated in the $yz$-plane of a different position in the $x$-direction. The  curves correspond to  position $x/L$ $=$ $0.1$, $0.55$ and $1$ (from the lighter to the darker grayscale).
%The lower curves correspond to  position $x/L=0.1$, and the upper correspond] to $x/L \approx 1$. The intermediary curves are separated in the $x$ coordinate by $\Delta x/L = 0.45$. 
The power spectrum were averaged in time from $t=3L/v_0$ to $t=4L/v_0$.}
\label{fig:aiii-ps_evol}
\end{figure}

The scales in which the CR pressure induces turbulence are shown to be concentrated in the typical scale of the density fluctuations, as expected from the BJL09 theory, at least while the density field spectrum is not modified by the precursor turbulence itself. Figure~\ref{fig:aiii-ps_evol} compares the power spectrum of the density and velocity field in different $yz$-planes inside the precursor for model AIII. The peaks of both power spectrum are approximately at the same wavenumbers, going to larger scales along $x$.

Therefore, it is clear that turbulence has not enough time to fully develop before 
the material reaches the shock, in the sense that the turnover time of the largest eddies inside the precursor is larger or comparable to the time it takes for the material to cross the precursor (however this time is not fixed, as the turbulent velocity increases along the precursor).

\begin{figure}
\centering
\includegraphics[width=.7\linewidth, trim= 0cm 0cm 0cm 0cm, clip=true,angle=270]{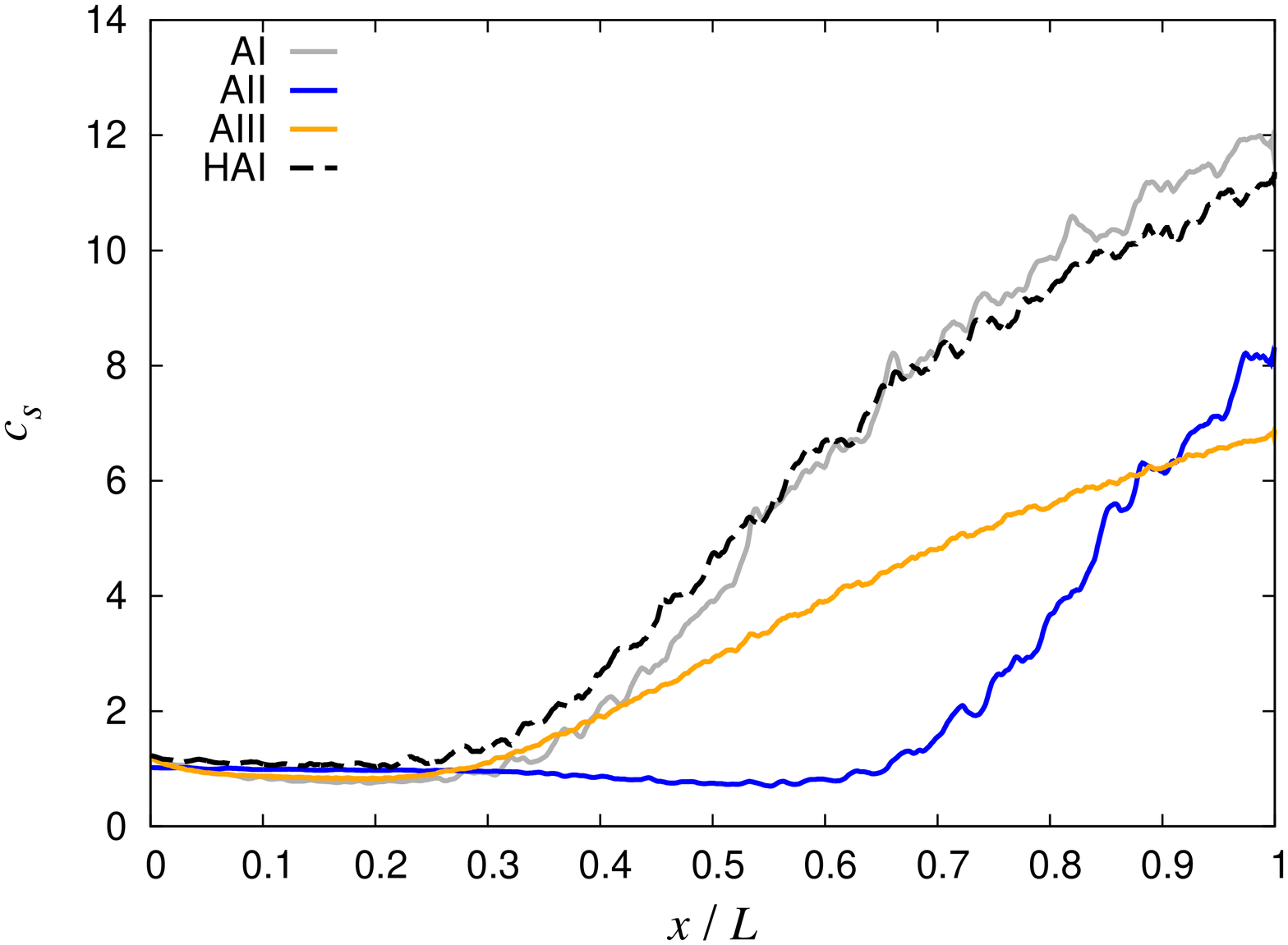} \\
\includegraphics[width=.7\linewidth, trim= 0cm 0cm 0cm 0cm, clip=true,angle=270]{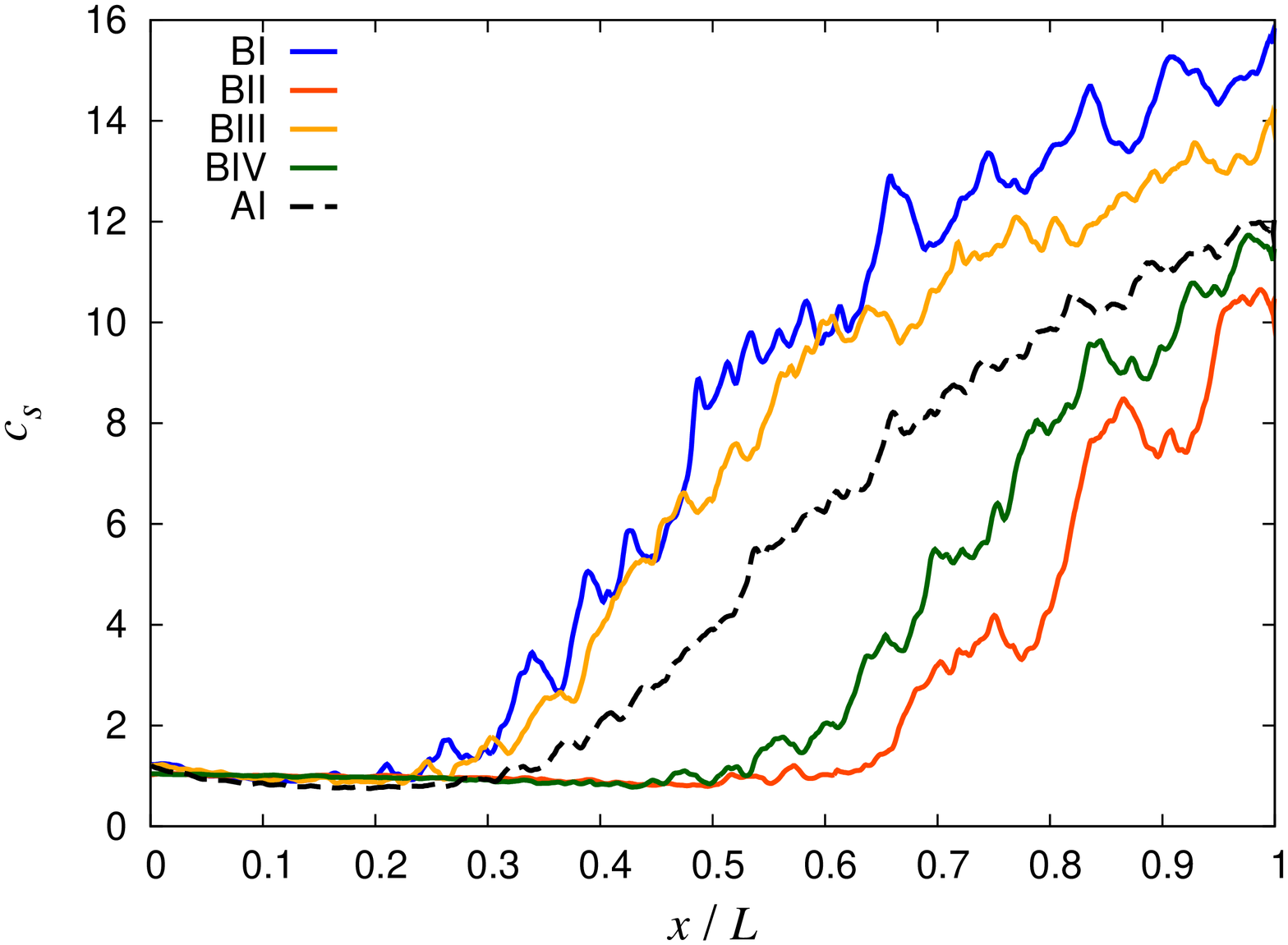}
\caption{Profile of the sound speed $\langle c_s \rangle = \langle \sqrt{\gamma P / \rho} \rangle$
for models A ({\it top}) and models B ({\it bottom}). 
The profiles were averaged in time from $t=3L/v_0$ to $t=4L/v_0$.}
\label{fig:csnd_profile}
\end{figure}

It could be thought that reducing the density structures size could give more time to the turbulence cascade to develop, increasing the efficiency of the small-scale dynamo. In fact, for model AIII the power spectrum of the magnetic field is closer to the solenoidal velocity field at the smaller scales, although it happens already for scales where the numerical dissipation is important;  therefore, we can not exclude the possibility that this would be different for  different grid resolutions (or different numerical dissipations). However, our model AIII show that the turbulent velocity increases more slowly and saturates at lower values in the precursor when comparing with model AI (see Figure~\ref{fig:decos}), at the same time that the maintenance of the density structures is degraded, despite that the turbulent heating of the gas along the precursor decays, increasing its compressibility. Figure~\ref{fig:csnd_profile} shows the comparison of the sound speed profiles between the models studied. 

\begin{figure}
\centering
\includegraphics[width=.7\linewidth, trim= 0cm 0cm 0cm 0cm, clip=true,angle=270]{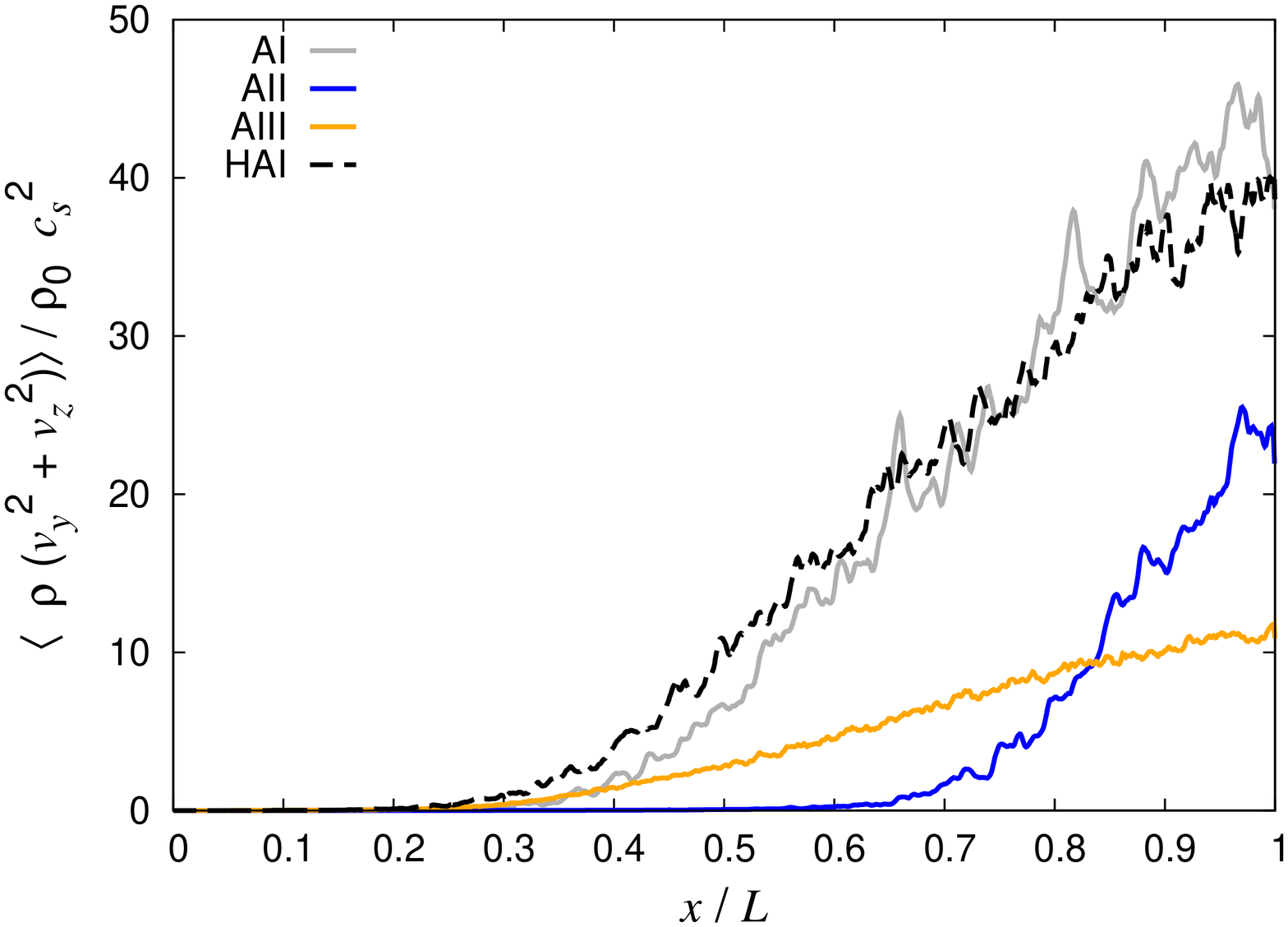} \\
\includegraphics[width=.7\linewidth, trim= 0cm 0cm 0cm 0cm, clip=true,angle=270]{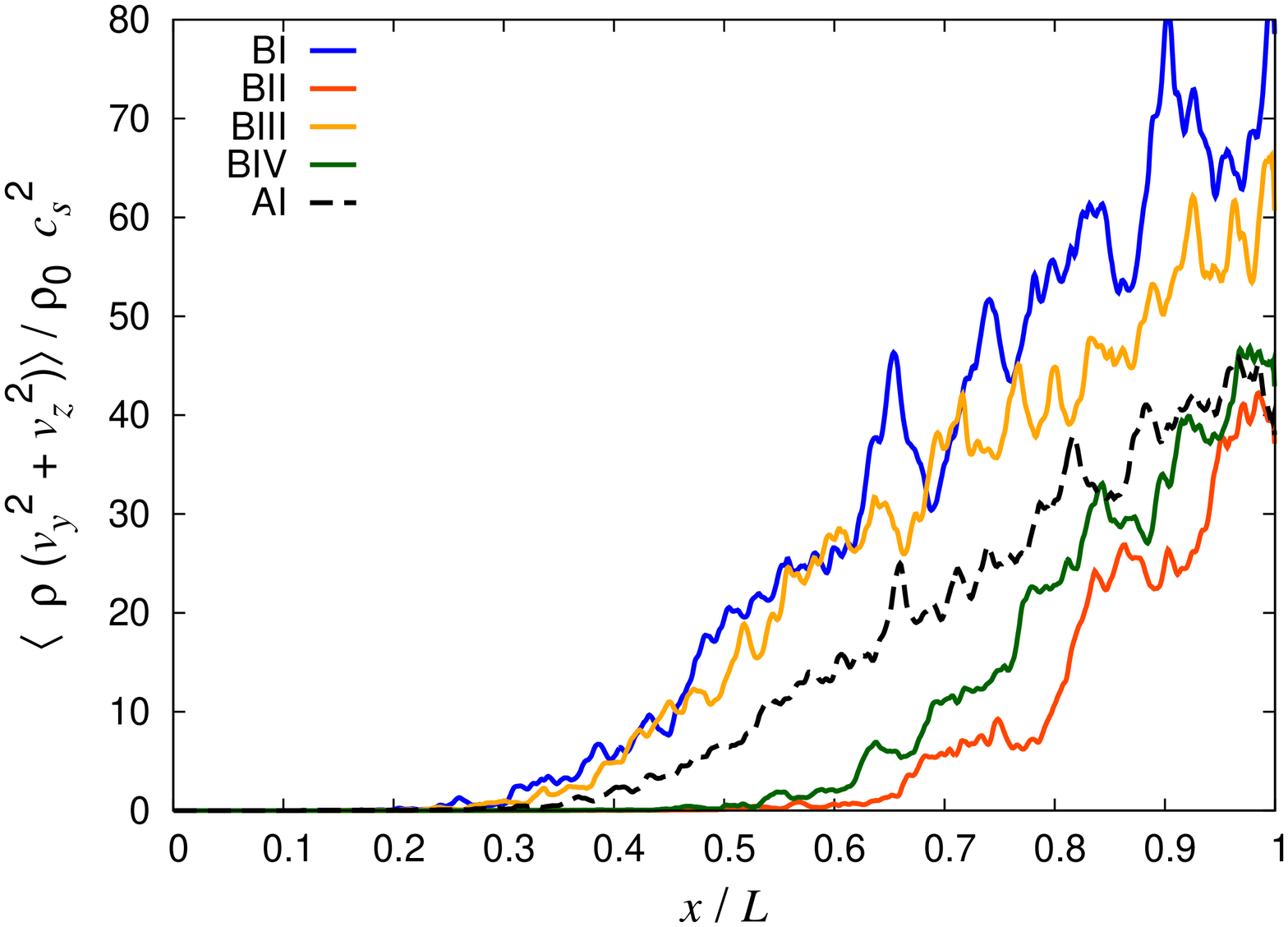}
\caption{Profile of the partial kinetic energy $\langle 0.5\,\rho (v_y^2 + v_z^2) \rangle$
for models A ({\it top}) and models B ({\it bottom}). The profiles were averaged in time from $t=3L/v_0$ to $t=4L/v_0$.}
\label{fig:ekin_v-yz}
\end{figure}

This result shows that the injection of turbulent energy in the system by the CR pressure is comparatively larger when the density structures have larger sizes. Figure~\ref{fig:ekin_v-yz} compares the profile of ``partial'' kinetic energy between models, calculated taking into account only the components $y$ and $z$ of the velocity field.

Therefore, we can conclude that the magnetic energy amplification is larger for larger values of the initial density contrast and larger $\lambda_{\rho}/D_{CR}$. However, this efficiency can not grow indefinitely, because the sound speed of the gas in the precursor increases as well, decreasing the compressibility of the gas (and consequently the maintenance of the density structures) and the effective Mach number of the shock, reducing the efficiency of the CR acceleration in the shock (not considered in our formulation).

The fact that the magnetic energy amplification is less efficient as  smaller the values of the injection scale $\lambda_{\rho}$ are can be understood in terms of the  mixing time of turbulence $t_{\rm mix}$  and the generation time $t_{\rm gen}$ of the turbulence-induced density structures. The former  can be estimated as $t_{\rm mix}$ $\sim$ $\lambda_{\rho}^{2}/\nu_{\rm turb}$, where $\nu_{\rm turb}$ is the eddy diffusivity, that can be estimated as $\lambda_{\rho}v_{\rm s}$, then $t_{\rm mix}$ $\sim$ $\lambda_{\rho}/v_{\rm s}$. $t_{\rm gen}$ can be estimated as $\lambda_{\rho}/v_{\rm p}$. Because $v_{\rm s} > v_{\rm p}$ then  $t_{\rm mix} < t_{\rm gen}$. For smaller values of $\lambda_{\rho}$, this condition saturates sooner, reducing the magnetic field amplification efficiency. Turbulent heating is not responsible of saturating the density structures as can initially be thought.
% because the plasma becomes more incompressible.

\subsection{Resolution effects}

The fact that a considerable part of the magnetic energy generated  occurs at the smallest scales of the simulations makes the numerical resolution important for the results. In fact, not only the numerical dissipation of the magnetic field but also the numerical dissipation  of the velocity (and also density) structures in these small scales cause a reduction of the magnetic energy amplification. 
%The Model HAI (see Table 1) has very similar parameters  to model AI, with the exception of the increased grid resolution. 

The model HAI presents a faster growth rate of vorticity with distance, compared to model AI; but the former presents a saturation value while the last increases continuously until the shock position (see Figure~\ref{fig:omg-a1}). The final magnetic energy density for  model HAI is approximately twice the value for the model AI ($B/B_0 \approx 200$ for model HAI and $\approx 100$ for model AI, according  to Figure~\ref{fig:rho-rms}). However, the profiles for the fluctuations of the solenoidal and potential velocity, and for the density, do not differ much between them, as shown in Figure~\ref{fig:decos}, which have the values dominated by the fluctuations at the larger scales (at the scales $\sim \lambda_{\rho}$). The power spectrum of the magnetic field shows the larger power for model HAI at the smaller scales (see Figure~\ref{fig-ps1}). 

These results show that the magnetic energy amplification obtained in the present simulations can be  underestimated, and should be considered as a lower limit.

\subsection{Synthetic versus MHD turbulence density fluctuations}

\begin{figure}
\begin{center}
\begin{tabular}{c c}
\includegraphics[width=.7\linewidth, trim= 0cm 0cm 0cm 0cm, clip=true,angle=270]{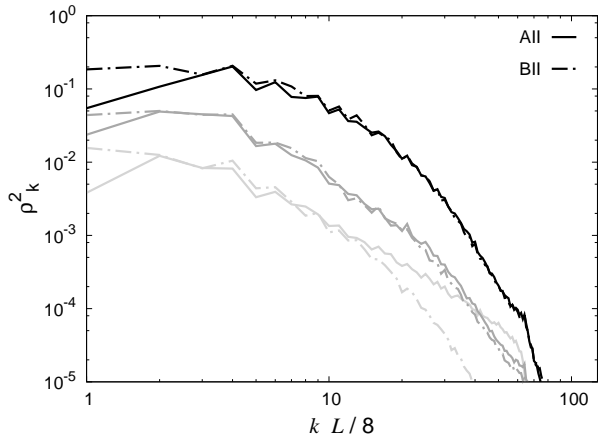}
\end{tabular}
\end{center}
\caption{Power spectrum of the density field $\rho$  for models AII (continuous lines) and for model BII (dashed lines). Each curve represents the power spectrum calculated in the $yz$-plane of a different position in the $x$-direction. The  curves correspond to  position $x/L$ $=$ $0.1$, $0.55$ and $1$ (from the lighter to the darker grayscale).%Each curve represents the power spectrum calculated in the $yz$-plane of a different position in the $x$-direction. The lower curves are for position $x/L=0.1$, and the upper correspond to $x/L \approx 1$.
The intermediary curves are separated in the $x$ coordinate by $\Delta x/L = 0.45$.
The power spectra were averaged in time from $t=3L/v_0$ to $t=4L/v_0$.}
\label{fig:dens_AB_evol}
\end{figure}

The comparison  between the results from models A and B shows that the statistical details of the density fluctuations have little influence on the final magnetic field amplification. In the setup employed in this study, the density structures inflowing in the precursor  lose their initial characteristics in a short distance, being dominated by the density structures induced by the compressional turbulence. In order to  illustrate this point, Figure~\ref{fig:dens_AB_evol} compares the evolution of the density spectrum along the precursor for models AII and BII, which have similar initial values of $\delta \rho_{rms} / \rho_0$ and $\lambda_{\rho}/D_{CR}$. We see that after $x/L \approx 0.5$ both spectrum present similar shapes.

However, it should be observed that the initial density spectrum of models A and B differ on scales larger than $\lambda_{\rho}$, with more power in models B. The same occurs for the magnetic field spectrum close to the shock position, as can be seen comparing the models in Figure~\ref{fig-ps1}. Models with similar initial values of $\delta \rho_{rms} / \rho_0$ and $\lambda_{\rho}/D_{CR}$ but slightly different spectral slope (models AI and BI, AII and BII) present similar values of magnetic field amplification (see Fig.~\ref{fig:rho-rms}).

\subsection{Dependence on the shock Mach number}

We ran an additional model using the same parameters as model AI (see Table~\ref{tab:amodels}), but with  ${\cal M}_{\rm s} = 400$. The resulting CR pressure is increased approximately by a factor $\sim$ 16, and consequently the turbulent energy is greater in this system. However,  there is no enough time for the magnetic field to grow further, and the final magnetic field is almost the same as in the original case with ${\cal M}_{\rm s} = 100$. 

Due to our self-regulated formulation, we do not expect great differences in the magnetic field amplification with the shock Mach number (we are not considering the dependences that the particle acceleration efficiency might have with ${\cal M}_{\rm s}$); although a very different value might produce differences in the magnetic field amplification.  However we are interested in strong non-relativistic Galactic shocks, in which the Mach numbers do not differ significantly.

\subsection{Limitations of the present study}\label{limitations}

The limited domain size in the direction parallel to the shock propagation might  reduce the magnetic field amplification. We considered the CR pressure acting only inside the scale $L = D_{CR}$, which means that the effects of the force at the distances where it decayed by a factor smaller than $\approx 3$ is neglected. In order to test this we made a run with the initial conditions given by model AI but doubling the box dimension in the $x$-direction, i.e. $L = 2 D_{CR}$, and resolution $2048\times128\times128$. The final magnetic field in this case is approximately twice the value obtain for model AI (not shown). 

Less important, if the computational box in the  perpendicular directions ($y$ and $z$) were larger, some small amplitude density fluctuations with scales larger than $\lambda_{\rho}$ could develop, leading to additional magnetic field amplification.

Also the presence of cooling could increase the generation of higher density fluctuations. Our models with higher initial density contrast have also higher magnetic field amplifications. However, the presence of  cooling was considered in \citet{downes14} (see next section), and they did not detected a substantial increase on the magnetic energy amplification.

As mentioned in Section~\ref{sec:data},  we do not consider the turbulent fluctuations of  velocity and magnetic field in the initial conditions of models B; this can be justified taking into account that the real ISM density structures are formed in the presence of cooling; therefore the density fluctuations are not totally dependent on the presence of magneto-acoustic modes, and can be present even after the decays of velocity and magnetic turbulence fluctuations (entropy modes). It is interesting to note that the velocity and magnetic fluctuations present from the  MHD turbulence simulations have larger amplitudes than those generated by the CRs pressure. Hence, if an initial turbulent magnetic field component is considered for the initial state (in the upstream material), the  magnetic field amplification close to the shock will be lower compared to our results.

Finally, we do not consider the CR pressure $P_{\rm CR}$ in the total pressure $P$ (see Eqs.~(3)-(7)). Adding $P_{\rm CR}$ to the total pressure might lead to a more incompressible plasma, hence the density fluctuations might saturate faster, possibly reducing the magnetic field amplification.   

\subsection{Comparison with previous studies}

The first numerical study exploring the magnetic field amplification in the shock upstream via MHD turbulence induced by  CR shock precursor was presented in \citet{drury12}. Employing 2D simulations for a shock propagating perpendicular to  the initially uniform magnetic field, they studied the magnetic field profile inside the precursor for different amplitudes of the initial density fluctuations, and also for different values of the initial magnetic field intensity. They found that the final magnetic energy amplification factor
depends  little on the amplitude of the initial density fluctuations. It was interpreted as  resulting from non-linear effects of turbulence, which increases the density contrast along the precursor and then reduces the influence of the initial values. They also found that the magnetic field amplification does not depend on its initial value, as it could be expected if the magnetic energy is far below the equipartition.

Later on, \citet{downes14} investigated the influence of the initial magnetic field orientation, the presence of cooling, and more importantly, they performed three-dimensional numerical simulations. They showed that the resulting magnetic field profile in the 3D setup is very similar to that of the 2D one,
with a slightly smaller magnetic field amplification in the former case. The magnetic field power spectrum are clearly different in the 2D and 3D cases as expected: in the 3D setup the magnetic field structures 
have smaller sizes; this difference arises because the velocity cascade of 2D turbulence proceed as an inverse cascade (as in hydrodynamic turbulence), 2D turbulence is not appropriate for modelling MHD turbulence. 
Concerning the initial magnetic field orientation, they found that larger amplification is produced in the perpendicular case. However, this difference in amplification with the parallel case can be explained, as showed in the paper,  to be produced by  compression. Finally, they showed that for the density scales they considered (representative of the ISM) cooling has little impact on their 2D simulations,  because the cooling time is longer than the time the gas takes to cross the precursor. 

%Nonetheless, considering a denser ISM (e.g., a molecular cloud), cooling might increase the magnetic field amplification.

Similar to the study in \citet{downes14}, the magnetic field amplification in our simulations does not  converge; we also attribute this to the fact that amplification along the precursor is still occurring  
at scales below the scales that can be resolved in our simulations. In fact, the solenoidal velocity is also continuously 
increasing along the precursor, although after passing the middle of the box (in the $x$ direction) the power spectrum shape remains approximately unchanged 
until the end of the domain (see Fig.~\ref{fig:aiii-ps_evol}). We have no evidences that the small-scale dynamo in any of our simulations 
enters  the linear phase after equipartition is already achieved at some scale. 
If the dynamo is still in the exponential growth phase, it can 
help to explain the similarities between the 2D and 3D amplifications found in \citet{downes14}, because the turbulence cascade is still not regulating the amplification of the field in the 3D case.%\footnote{Differences are naturally expected, because there is no such 2D dynamo.}. 
We have inferred that the turn-over time of the turbulence outer scale in our setups is not short compared with the time these eddies remain inside the precursor.

When studying  the role of  cooling, \citet{downes14} focused on their 2D models. 
It is expected that in the 3D case, where the forward cascade quickly heats the gas
reducing its compressibility and then the generation of density structures, 
the cooling to have a stronger impact. Another important point is related to the generation of CRs in the shock. The acceleration process is though to be sensitive to the sonic Mach number of the shock. Therefore the turbulence heated gas in the precursor also has the effect of reducing the CR pressure, which in turn reduces the magnetic field amplification in the precursor. However, the influence of the precursor on the shock evolution itself is not considered in our simplified model, which requires a much more involved modelling of the problem.

A  difference between our approach and that in \citet{drury12} and \citet{downes14} 
is in the model of the CR pressure in the precursor. In our approach, the CR pressure is considered to decay exponentially (instead of linearly) from the shock front, and consequently the CR force also decays exponentially%\footnote{\bf However, as we are unable to calculate the pressure distribution properly, both approaches are equally speculative.}. 
Although the self-consistent profile of the CR pressure is difficult to calculate --- as it depends on the energy of the CR rays considered and their diffusivity which in turn is anisotropic and depends on the physics of the upstream flow  (such as the magnetic fields fluctuations, the velocity profile, etc.) --- we just considered the solution for the CR energy in the diffusion approximation for a fixed value of the diffusivity, and without considering the influence of the CR pressure on the upstream flow.

\begin{figure}
\centering
\begin{tabular}{c}
\includegraphics[width=.6\linewidth, trim= 0cm 0cm 0cm 0cm, clip=true,angle=270]{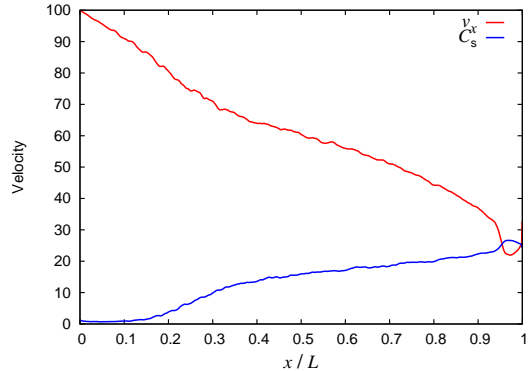}
\end{tabular}
\caption{Profiles of the flow velocity and of the sound speed $c_{\rm s} = \sqrt{\gamma p / \rho}$.
The simulation uses a CR force in of the form $F_{\rm CR} = -\eta \rho_0 v_{0}^2$.}
\label{fig-cs-comparison}
\end{figure}

Another difference in the CR pressure modeling, and probably the most relevant, 
is the use of the flow variables close to the shock in order to obtain the ram pressure of the flow at the shock. The CR force amplitude is updated dynamically 
in our simulations, accordingly to the kinetic energy of the upstream flow available at the shock. The previous works considered a fixed force
$F_{\rm CR} = - \eta \rho_0 v_0^2 / D_{CR}$, which
can lead to non-physical situations (no shock) if the induced turbulence in the precursor increases the sound speed in the gas above the upstream velocity, 
making the problem ill-posed; this can be seen in Fig.~\ref{fig-cs-comparison}, where the fluid velocity versus sound speed profiles are shown for a simulation with a constant CR force. The plasma reaches temperatures so high that the flow velocity gets lower than $c_{\rm s}$ (no shock) and then no Fermi acceleration would be operating.

\begin{figure}
\begin{centering}
\includegraphics[width=.7\linewidth, trim= 0cm 0cm 0cm 0cm, clip=true,angle=270]{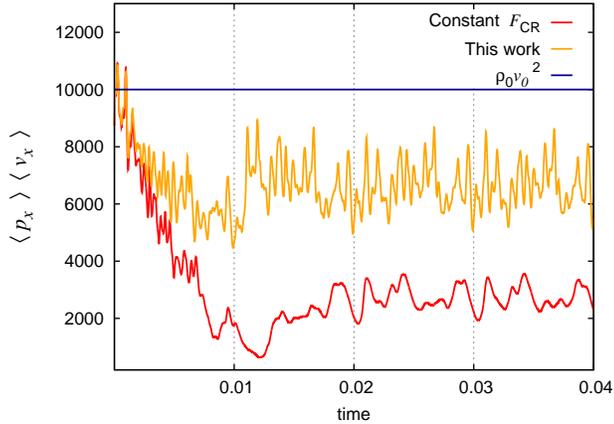}
\end{centering}
\caption{Mean flow  evolution of $\left\langle p_x \right\rangle\left\langle v_x \right\rangle$ at the right boundary, during the whole integration time. Each time period of $0.01$ is delimited by a dashed vertical line. 
Both simulations have synthetic turbulence initial conditions and the  same numerical set-up but the force formulation. The constant force corresponds to a force of the form 
$F_{\rm CR} = -\eta \rho_0 v_{0}^2 /D_{CR}$ \citep[e.g.,][]{drury12,downes14}, and in our set-up 
$F_{\rm CR}$ is given by Eq.~(\ref{force}). 
The blue line is the constant ram pressure used in  the constant force formulation.}
\label{fig-forces-comparison}
\end{figure}

Our formulation imposes a more stringent limit on the efficiency of the 
magnetic field amplification mechanism, which we believe to be 
more realistic: 
a stronger force  reduces more the pre-shock velocity flow, at the same time 
more turbulence is generated; 
this reduces the upstream flow ram pressure and 
 hence the CR acceleration becomes less efficient, reducing the CR pressure force. 
The system is self-regulated. 
From Fig.~\ref{fig-forces-comparison},  can be seen  that in the case of a constant force used in the previous 
works the flow ram pressure at the boundary is less than one third the value used $\rho_{0} v_{0}^{2}$ for the 
CR force calculation, that is, the amplitude considered for the CR force is overestimated.

% %%==============================================================================
% %%

\section{Summary and conclusions}
\label{sec:conclusions}
In this work we investigated turbulence-induced magnetic field amplification in shock 
precursors under the BJL09 model. We used MHD simulations to quantify the problem, 
modelling the CR pressure as a function of the flow ram pressure. We considered models with different 
density inhomogeneities in the precursor, parametrized the density contrast 
$\delta \rho_{rms}/\rho_0$ and by the relative density structures scales $\lambda_{\rho} / D_{CR}$.

In agreement to BJL09, we have found that:

\begin{itemize}
		
\item The interaction of density fluctuations with the CR precursor produces vorticity. 
Our calculations show that vorticity is generated at a faster rate for higher density contrast and 
for smaller density structure scales;

\item The solenoidal velocity 
fluctuations dominate over the compressible ones in the  turbulence induced at the precursor. 
Both components achieve roughly $\lesssim$ 10\% of the upstream 
velocity and their evolution depends strongly on the initial density contrasts and density structures sizes;

\item The scales in which the CR pressure induces turbulence in the precursor are close 
to the scale of the density inhomogeneities. Hence, the injection of turbulent 
energy in the precursor is larger when the ISM density structures have larger sizes.

\end{itemize}

The final magnetic energy depends strongly on  $\delta \rho_{rms}/\rho_0$, but it is approximately insensitive 
to the values of $\lambda_{\rho} / D_{CR}$ considered here. 
In our formulation, the amplification factors of the magnetic field
due to the gas compression are relatively small ($\sim 1.3$), 
being most of the magnetic energy growth due to the small-scale dynamo.

The amplified magnetic fields we obtained in this work are still concentrated 
at the small scales, indicating that the dynamo do not have enough time to reach 
the linear phase. The results then depend on the numerical resolution employed, 
and the magnetic energy amplifications obtained should be considered as  lower limits.
The maximum amplification factor for the magnetic field we obtained is $\sim$ 14, for model HAI. 
When considering a larger domain, the amplification factor duplicates (see Sect.~\ref{limitations}), 
therefore the amplified magnetic field that we obtained is $B_{\rm cgs} \geq 100$~$\mu$G.
This lower limit value is in agreement with the inferred values of the magnetic fields in Cassiopeia A ($\sim$ 100 $\mu$G) and Tycho's SNRs ($\sim$ 300 $\mu$G) \citep{vink03,cassam07}. 

We found that, under the setup considered, the statistical details of the density fluctuations
(power spectrum slope and anisotropies in the density structures) 
have little influence on the final magnetic field amplification. Therefore, for the same 
$\delta\rho_{rms}/\rho_0$ and $\lambda_{\rho}/D_{CR}$, no significant differences were found between density fields built
synthetically or obtained from MHD turbulence.

\section*{Acknowledgments}
A. L. acknowledges the NSF grant AST 1212096 and Center for Magnetic Self Organization (CMSO) as well as a distinguished visitor PVE/CAPES appointment at the Physics Graduate Program of the Federal University of Rio Grande do Norte and thanks the INCT INEspao. RSL acknowledges support from a grant of the Brazilian Agency FAPESP (2013/15115-8). This work has made use of the computing facilities of the Laboratory of Astroinformatics (IAG/USP, NAT/Unicsul), whose purchase was made possible by the Brazilian agency FAPESP (grant 2009/54006-4) and the INCT-A. RSL and MVdV thank the hospitality of the Astronomy department of the University of Wisconsin/Madison and of the Departamento de F\'isica Te\'orica e Experimental of the Universidade Federal do Rio Grande do Norte, where part of this work was developed. 
 
%%%%%%%%%%%%%%%%%%%%%%%%%%%%%%%%%%%%%%%%%%%%%%
%%==============================================================================
%%


\begin{thebibliography}{}
\label{bibliography}
\bibitem[Armstrong, Rickett \& Spangleret (1995)]{armstrong95} Armstrong, J.~W., 
Rickett, B.~J., \& Spangler, S.~R.\ 1995, \apj, 443, 209 
\bibitem[Axford, Leer \& McKezie (1982)]{Axford82}Axford, W.~I., Leer, E., \& McKezie, J.~F. 1982, A\&A, 111, 317
%\bibitem[Ballesteros-Paredes (2006)]{Ballesteros-Paredes06}Ballesteros-Paredes, J. 2006, MNRAS, 372, 443
\bibitem[Bell (1978)]{Bell}Bell, A.~R. 1978, MNRAS, 182, 147
\bibitem[Bell (2004)]{Bell04}Bell, A.~R. 2004, MNRAS, 353, 550
%\bibitem[Beresnyak, Lazarian \& Cho (2005)]{Beresnyak05}Beresnyak, A., Lazarian, A. \& Cho, J. 2005, ApJ, 624, L93
\bibitem[Beresnyak, Jones \& Lazarian (2009)]{Beresnyak09}Beresnyak, A., Jones, T.~W. \& Lazarian, A. 2009, ApJ, 707, 1541
\bibitem[Beresnyak \& Li (2014)]{beresnyak14}Beresnyak, A. \& Li, H. 2014, \apj, 
788, 107
\bibitem[Beresnyak \& Lazarian(2015)]{2015ASSL..407..163B} Beresnyak, A., \& Lazarian, A.\ 2015, Magnetic Fields in Diffuse Media, 407, 163
\bibitem[Blassi (2013)]{blasi2013}Blasi, P. 2013, A\&A Rev., 21, 70
\bibitem[Br\"{u}ggen (2013)]{bruggen13}Br\"{u}ggen M., 2013, MNRAS, 436, 294
\bibitem[Burkhart et al. (2009)]{burkhart09}Burkhart, B., Falceta-Gon\c{c}alves, D., Kowal, G. \& Lazarian, A. 2009, ApJ, 693, 250
\bibitem[Burkhart et al.(2012)]{2012ApJ...749..145B} Burkhart, B., Lazarian, A., \& Gaensler, B.~M.\ 2012, \apj, 749, 145
\bibitem[Cassam-Chena\"{i} et al. (2007)]{cassam07}Cassam-Chena\"{i}, G., Hughes, J.P., Ballet, J.  \&  Decourchelle, A. 2007, \apj, 665, 315
\bibitem[Chepurnov \& Lazarian (2010)]{chepurnov10}Chepurnov, A. \& Lazarian, A. 2010, \apj, 710, 853 
\bibitem[Chepurnov et al. (2010)]{chepurnov10b}Chepurnov, A., Lazarian, A. Stanimirovi\'{c}, S., Heiles, C., Peek, J.~E.~G. 2010, \apj, 714, 1398
\bibitem[Cho et. al(2009)]{cho09}Cho, J., Vishniac, E., Beresnyak, A., Lazarian, A., Rye, D. 2009, ApJ 693, 1449
\bibitem[Crutcher et al. (1999)]{crutcher99} Crutcher, R.~M., Roberts, D.~A., Troland, T.~H. \& Goss, W.~M. 1999, ApJ, 515, 275
%\bibitem[de Gouveia dal Pino \& Lazarian(2005)]{2005A&A...441..845D} de Gouveia dal Pino, E.~M., \& Lazarian, A.\ 2005, \aap, 441, 845
\bibitem[Diamond \& Malkov (2007)]{Diamond2007}Diamond, P.~H., \& Malkov, M.~A. 2007 ApJ, 654, 252
\bibitem[Downes \& Drury (2014)]{downes14} Downes, T.~P.  Drury, L.~O.'C 2014, MNRAS, 444, 365
\bibitem[Draine \& Lazarian (1998)]{Draine98}Draine, B.~T., \& Lazarian, A. 1998, ApJL, 494, L19
\bibitem[Drury (1984)]{drury84}Drury, L.~O'C. 1984, Adv. Space Res., 4, 185
\bibitem[Drury \& Downes (2012)]{drury12} Drury, L.~O.'C \& Downes, T.~P. 2012, MNRAS, 427, 2308
\bibitem[Elmegreen \& Scalo (2004)]{elmegr2004}Elmegreen, B.~G., \& Scalo, J. 2004, ARA\&A, 42, 211
\bibitem[Falceta-Gon\c{c}alves et al. (2014)]{falceta14}Falceta-Gon\c{c}alves, D., Kowal, G., Falgarone, E. \& Chian, A.~L. 2014, Nonlinear Processes in Geophysics, 21, 587
\bibitem[Falgarone et al. (2008)]{falgarone08} Falgarone, E., Troland, T.~H., Crutcher, R.~M. \& Paubert, G. 2008, A\&A, 487, 247
\bibitem[Federrath et al. (2010)]{federrath10} Federrath, C., Roman-Duval , J., Klessen, R.~S., Schmidt, W. \& Mac Low, M.~M. 2010, A\&A, 512, 81
\bibitem[Giacalone \& Jokipii (1999)]{giacalone99}Giacalone, J. \& Jokipii, J.~R. 1999, ApJ, 520, 204
\bibitem[Giacalone \& Jokipii (2007)]{giacalone07}Giacalone, J. \& Jokipii, J.~R. 2007, ApJ, 663, L41
\bibitem[Ginzburg \& Syrovatskii (1964)]{ginzb}Ginzburg, V. \& Syrovatskii, S. 1964, Origin of Cosmic Rays, Pergamon Press, NY
\bibitem[Gou at al. (2012)]{gou12} Guo, F., Li, S., Li, H., Giacalone, J., Jokipii J.~R. \& Li, D. 2012, ApJ, 747, 98
\bibitem[Haverkorn et al. (2008)]{haverkorn08}Haverkorn, M., Brown, J.~C., Gaensler, B.~M., \& McClure-Griffiths, N.~M. 2008, \apj, 680, 362
\bibitem[Hill et al.(2008)]{2008ApJ...686..363H} Hill, A.~S., Benjamin, R.~A., Kowal, G., et al.\ 2008, \apj, 686, 363
%\bibitem[Joung \& Mac Low (2006)]{joung06}Joung, M.~K.~R., \& Mac Low, M.~M. 2006, ApJ, 653, 1266
\bibitem[Kowal et al. (2007)]{kowal07} Kowal, G., Lazarian A., \& Beresnyak, A. 2007, ApJ, 658, 423
\bibitem[Kowal et al.(2012)]{2012PhRvL.108x1102K} Kowal, G., de Gouveia Dal Pino, E.~M., \& Lazarian, A.\ 2012, Physical Review Letters, 108, 241102
\bibitem[Lagage \& Cesarsky (1983)]{lagage83}Lagage, P.~O. \& Cesarsky, C.~J. 1983, A\&A. 125, 249L
%\bibitem[Lazarian \& Vishniac(1999)]{1999ApJ...517..700L} Lazarian, A., \& Vishniac, E.~T.\ 1999, \apj, 517, 700
\bibitem[Lazarian \& Opher (2009)]{laz_oph2009}Lazarian, A. \& Opher, M. 2009, ApJ, 703, 8
\bibitem[Lazarian \& Gaensler (2012)]{lazarian12}Lazarian, A., \& Gaensler, B.~M.\ 2012, \apj, 749, 145
%\bibitem[Le\~{a}o et al. (2009)]{leao09}Le\~{a}o, M.~R.~M., de Gouveia Dal Pino, E.~M., Falceta-Gon\c{c}lves, D., Melioli, C., \& Geraissate, F.~G. 2009, MNRAS, 394, 157
\bibitem[Mac Low \& Klessen(2004)]{mac04} Mac Low, M.~M. \& Klessen, R.~S. 2004, Rev. of Mod. Phys., 76, 125
%\bibitem[Mac Low (2009)]{mac09}Mac Low, M.~M. 2009, Revista Mexicana de Astronomia y Astrofisica Conference Series, 36, 121
\bibitem[Malkov \& Diamond (2009)]{malkov09}Malkov, M.~A. \& Diamond, P.~H. 2009, ApJ, 642, 244 
\bibitem[Malkov \& Drury (2001)]{malkov2001}Malkov, M.~A. \& Drury, L.~O'C. 2001, Rep. Prog. Phys. 64, 429
\bibitem[Markevitch et al. (2005)]{markevitch05}Markevitch M., Govoni F., Brunetti G. \& Jerius D. 2005, \apj, 627, 733
%\bibitem[McCray \& Snow (1979)]{McCray79}McCray, R., \& Snow, T.~P., Jr. 1979, ARA\&A, 17, 213
\bibitem[McKee \& Ostriker (2007)]{mckee2007} McKee, C. \& Ostriker, E. 2007, ARA\&A, 45, 565
\bibitem[Mignone et al. (2007)]{mignone07} Mignone, A., Bodo, G., Massaglia, S., et al. 2007, ApJSS, 170, 228
%\bibitem[Melioli \& de Gouveia Dal Pino (2006)]{melioli06}Melioli, C. \& de Gouveia Dal Pino, E.~M. 2006, A\&A, 445, L23
%\bibitem[Melioli et al. (2009)]{melioli09}Melioli, C.,Brighenti, F., D'Ercole, A. \& de Gouveia Dal Pino, E.~M. 2009, MNRAS, 399, 1089
\bibitem[Mizuno et al. (2014)]{mizuno14}Mizuno, Y., Pohl, M., Niemiec, J., Zhang, B., Nishikawa, K. \& Hardee, P.~E. 2014, MNRAS, 439, 3490
\bibitem[Schekochihin \& Cowley (2007)]{schekoch2007} Schekochihin, A. \& Cowley, S. 2007, in 
   {\it Magnetohydrodynamics - Historical evolution and trends},
   eds. by S. Molokov, R. Moreau, \& H. Moffatt
   (Berlin; Springer), p.~85. (astro-ph/0507686)
\bibitem[Schure et al. (2012)]{schure12}Schure, K.~M., Bell, A.~R., Drury, L.~O'C. \& Bykov, A.~M., 2012, Space Sci REv 173, 491
\bibitem[Uchiyama et al. (2007)]{uchiyama09}Uchiyama, Y., Aharonian, F.~A., Tanaka, T., et al. 2007, Nature, 449, 576
\bibitem[van Weeren et al. (2011)]{vanWeeren11}van Weeren, R.~J., Br\"{u}ggen M., R\"{o}ttgering H.~J.~A. \& Hoeft, M., 2011, \mnras, 418, 230
\bibitem[Vink \& Laming (2003)]{vink03}Vink, J., \& Laming, J.M. 2003, \apj, 584, 758
\bibitem[Vink (2012)]{vink12}Vink, J., 2012, A\&AR, 20, 49
%\bibitem[V\"olk, Drury \& McKenzie (1984)]{volk1984}V\"olk, H.J., Drury, L.O'C., \& McKenzie, J.~F. 1984, A\&A, 130, 19
\bibitem[Zeldovich et al. (1984)]{zeld84}Zeldovich, Y.~B., Ruzmaikin, A.~A., Molchanov S.~A., \& Sokoloff, D.~D. 1984, J Fluid Mech, 144, 1

\end{thebibliography}
\end{document}